\documentclass[12pt]{iopart}

\usepackage{iopams}  
\usepackage{amssymb}
\usepackage{graphics}
\usepackage{float}
\usepackage{subfigure}
\usepackage{color}
\usepackage{epsfig}
\usepackage[bookmarks=false]{hyperref}
\usepackage{xcolor}
\hypersetup{colorlinks=true,citecolor=blue,
linkcolor=blue,urlcolor=blue,pdfstartview=FitH,
bookmarksopen=true}
\usepackage{mathrsfs}
\usepackage{upgreek}

\def\redRR#1{\textcolor{black}{#1}}
\def\red#1{\textcolor{black}{#1}}
\newcommand{\ot}{\otimes}
\newcommand{\ket}[1]{|{#1}\rangle}
\newcommand{\bra}[1]{\langle{#1}|}
\newcommand{\bkt}[2]{\langle{#1}|{#2}\rangle}

\begin{document}

\title[AWVA for gravitational-wave detection]{
A potential third-generation gravitational-wave detector based on autocorrelative weak-value amplification}

\author{Jing-Hui Huang$^{1,2}$, Fei-Fan He$^{1}$, Xue-Ying Duan$^{3}$, Guang-Jun Wang$^{3}$, Xiang-Yun Hu$^{1,*}$ }
\address{$^{1}$ School of Geophysics and Geomatics, China University of Geosciences, Wuhan 430074, China 
 }
 \address{$^{2}$ 
Department of Physics and Centre for Research in Photonics, University of Ottawa, 25 Templeton Street, Ottawa, Ontario, Canada K1N 6N5
 }
 \address{$^{3}$ School of Automation, China University of Geosciences, 430074 Wuhan, China.
 }

\ead{xyhu@cug.edu.cn}

\vspace{10pt}
\begin{indented}
\item[]June 2023
\end{indented}

\begin{abstract}
Reducing noises and enhancing signal-to-noise ratios (SNRs) have become critical for designing third-generation gravitational-wave (GW) detectors with a GW strain of less than $10^{-23}$/$\rm \sqrt{Hz}$. In this paper, we propose a potential third-generation GW detector based on autocorrelative weak-value amplification (AWVA) for GW detection with a strain of \red{$h_g =$ $4 \times 10^{-25}$/$\rm \sqrt{Hz}$}. 
\red{In our scheme,}
a GW event \red{induces} a phase difference $\Delta \phi$ \red{by passing} through a  \red{ 11-bounce delay line, 10-km arm-length, zero-area Sagnac interferometer illuminated with a 1064-nm laser}. \red{Subsequently,} $\Delta \phi$ is amplified as the parameter of post-selection by choosing the appropriate pre-selected state and coupling strength in AWVA. 
In particular, we theoretically investigate the AWVA measurements for GW detection \red{within the frequency band of 200 Hz $\leq$ $f_g$ $\leq$ 800 Hz, considering} Gaussian noises with negative-decibel signal-to-noise ratios.
The peak response of the AWVA sensitivity $\kappa(f_g)$ occurs at frequency $f_{g, max}$ = \red{500} Hz, \red{which falls within the  frequency band of interest of the current third-generation GW detectors}. 
Our simulation results indicate that AWVA can demonstrate a measurable sensitivity of the autocorrelation coefficient $\Theta(f_g)$ within the frequency band of interest. \red{Moreover, the robustness of WVA shows promising potential in mitigating the effects of Gaussian noises.} 
\end{abstract}


\section{Introduction}
\label{sec:introduction}
The first detected event of gravitational-wave (GW), GW150914~\cite{PhysRevLett.116.061102} with a peak GW strain of $h_g \approx$ $10^{-21}$/$\rm \sqrt{Hz}$ has \red{not only} made a major impact on our understanding of  astrophysical implications of the detected sources~\cite{PhysRevX.9.031040,Abbott2020-2,Abbott_2020,PhysRevD.107.023511} \red{but also contributed significantly to advancements in optical metrology and sensing}~\cite{PhysRevLett.104.251102,RevModPhys.94.025001,PhysRevD.107.022003,doi:10.1063/5.0117605}.
Currently, confidently-detected GW events are observed by second-generation detectors based on extremely sensitive Michelson interferometers: 
GEO-HF in German~\cite{HLuck_2010}, 
Advanced Virgo in Italy~\cite{Accadia_2011},
KAGRA in Japan~\cite{Somiya_2012}
and Advanced LIGO in the USA~\cite{Abbott_2020}. 
The strain noise level of KAGRA, Advanced Virgo and Advanced LIGO can reach 0.3$\times 10^{-23}$/$\rm \sqrt{Hz}$.
Nevertheless, numerous works are studying the \red{possibility of a potential third-generation detector to reach a lower strain noise and a higher sensitivity}~\cite{Hild_2011,RevModPhys.86.121,PhysRevD.97.064031,Hall_2019,luck2020design,Maggiore_2020}. 
\red{
Thereinto, the Einstein Telescope (ET)~\cite{Maggiore_2020}, a proposed European ground-based GW detector of the third generation with a 10-km arm length, 
could be operating at a lower strain noise of $h_g \approx$ $4 \times 10^{-25}$/$\rm \sqrt{Hz}$. This value is derived from the Strain noise levels of ET at $f_g$ = 500 Hz, as shown in Fig.~\ref{Fig:SagnacPhaseAndNoise}(b).
}

One potential advanced \red{third-generation} GW detector may be realized by zero-area Sagnac interferometers~\cite{PhysRevLett.104.251102,PhysRevLett.76.3053,Grf_2014}. The zero-area Sagnac interferometer is modified from the conventional Sagnac interferometer~\cite{1913L} and is insensitive to rotations \red{while} remaining sensitive to time-dependent displacement in the two arms.
Sun $et \, al.$ have investigated the \red{possibility} of detecting a broad-band GW detector with a zero-area Sagnac interferometer. \red{When compared with the peak response of a Michelson interferometer,} they found that the zero-area Sagnac interferometer can obtain a more significant peak response with the same storage time and power on the beam splitter of the Michelson interferometer~\cite{PhysRevLett.76.3053}. 
Furthermore, Eberle $et \, al.$ experimentally showed that the sensitivity of such an interferometer can surpass the standard quantum limit by inputting the squeezed-light laser~\cite{PhysRevLett.104.251102}. Therefore, the Sagnac topology with other quantum-enhanced techniques \red{could} be competitive for future GW detectors.

\red{Currently}, weak measurement (WM) has  \red{demonstrated} its advantages in \red{estimating} ultra-small quantities both in theoretic analysis and experimental implementation~\cite{RevModPhys.86.307,Lundeen2011,Turek_2015,Li_2019,Huang_2020,Mori_2020,Li_2021,PhysRevA.97.033851,PhysRevA.105.013718,AWVA,Harraz2022} since it was proposed by Aharonov $et \, al.$ in 1988~\cite{AAV}. \red{The standard WM technique involves four steps}: pre-selections for the two-level system at state $\ket{\Phi_{i}}$ and the pointer at \red{state}  $\ket{\Psi_{i}}$, \red{either} in the time or frequency domain; weak interaction between the system and the pointer; \red{post-selection of the final state $\ket{\Phi_{f}}$ of the system} and read-out of the shifts of the pointer. By \red{appropriately} choosing the pre-selected state $\ket{\Phi_{i}}$, post-selected state $\ket{\Phi_{f}}$ and the coupling strength, the unknown quality can be amplified by the factor \red{of} weak-value $A_{w}:= {\bra{\Phi^{f}}\hat{A}\ket{\Phi^{i}}}/{\bkt{\Phi^{f}}{\Phi^{i}}}$, \red{which serves} as one parameter of the system~\cite{PhysRevA.97.033851,doi:10.1063/1.5019816,He:21} or the weak-coupling strength~\cite{PhysRevA.105.013718,PhysRevLett.111.033604,PhysRevA.103.032212}.
\red{It is worth noting} that the amplification $A_{w}$ is inversely proportional to the \red{probability} $|\bkt{\Phi^{f}}{\Phi^{i}}|^2$ of the \red{successful} post-selection. In other words, \red{a larger enhancement makes it more challenging to detect the final pointer due to the environmental disturbance and instrument instability that inevitably influence the measurements. } 
Therefore, our previous work proposed \red{the} autocorrelative weak-value amplification (AWVA) \red{technique and simulated the protocol under the influence of noises with negative-decibel signal-to-noise ratios (SNRs)~\cite{AWVA}}.
The robustness of AWVA \red{motivates us to} investigate GW detection with a strain of \red{$h_g =$ $4 \times 10^{-25}$/$\rm \sqrt{Hz}$} \red{in the presence of strain noises with negative SNRs.}

\red{In this paper, we proposed a potential third-generation GW detector utilizing the AWVA technique in the time domain, with the capability to detect GW with a strain of $h_g =$ $4 \times 10^{-25}$/$\rm \sqrt{Hz}$}. \red{Our approach involves} a GW with ``plus'' polarization passing through a \red{zero-area} Sagnac interferometer and introducing a phase difference $\Delta \phi$ in the optical arm. 
The \red{GW-induced} phase difference was then amplified as a parameter of post-selection in one weak measurement, while another simultaneous weak measurement without phase difference $\Delta \phi$ was \red{conducted to obtain} the autocorrelation coefficient $\Theta$.
We \red{performed simulations of the} AWVA measurement for detecting the GW with a strain of \red{$h_g =$ $4 \times 10^{-25}$/$\rm \sqrt{Hz}$} under the influence of Gaussian-white noises. \red{We considered various SNRs of noises and different sampling rates of detectors}. 
Our results indicated that AWVA can demonstrate a measurable sensitivity of the autocorrelation coefficient $\Theta$ and the robustness in AWVA has the potential to resist Gaussian noises \red{with negative SNRs}. 

The paper is organized as follows.
In Section 2, \red{we present the scheme of zero-area Sagnac interferometers and display the GW-induced phase shift in different sizes of the interferometers}.
In Section 3, we show a potential third-generation GW detector based on AWVA in the time domain.
In Section 4, the simulation results at the GW frequency band \red{200 Hz $\leq$ $f_g$ $\leq$ 800 } Hz are obtained under Gaussian noises with \red{varoius SNRs and different sampling rates of detectors}.
Finally, in Section 5, we give the conclusion about the work. 

\section{zero-area Sagnac interferometer}
\label{sec:zero-areaSagnac}
\begin{figure*}[t]
\centering
\resizebox{0.99\textwidth}{!}{%
\includegraphics{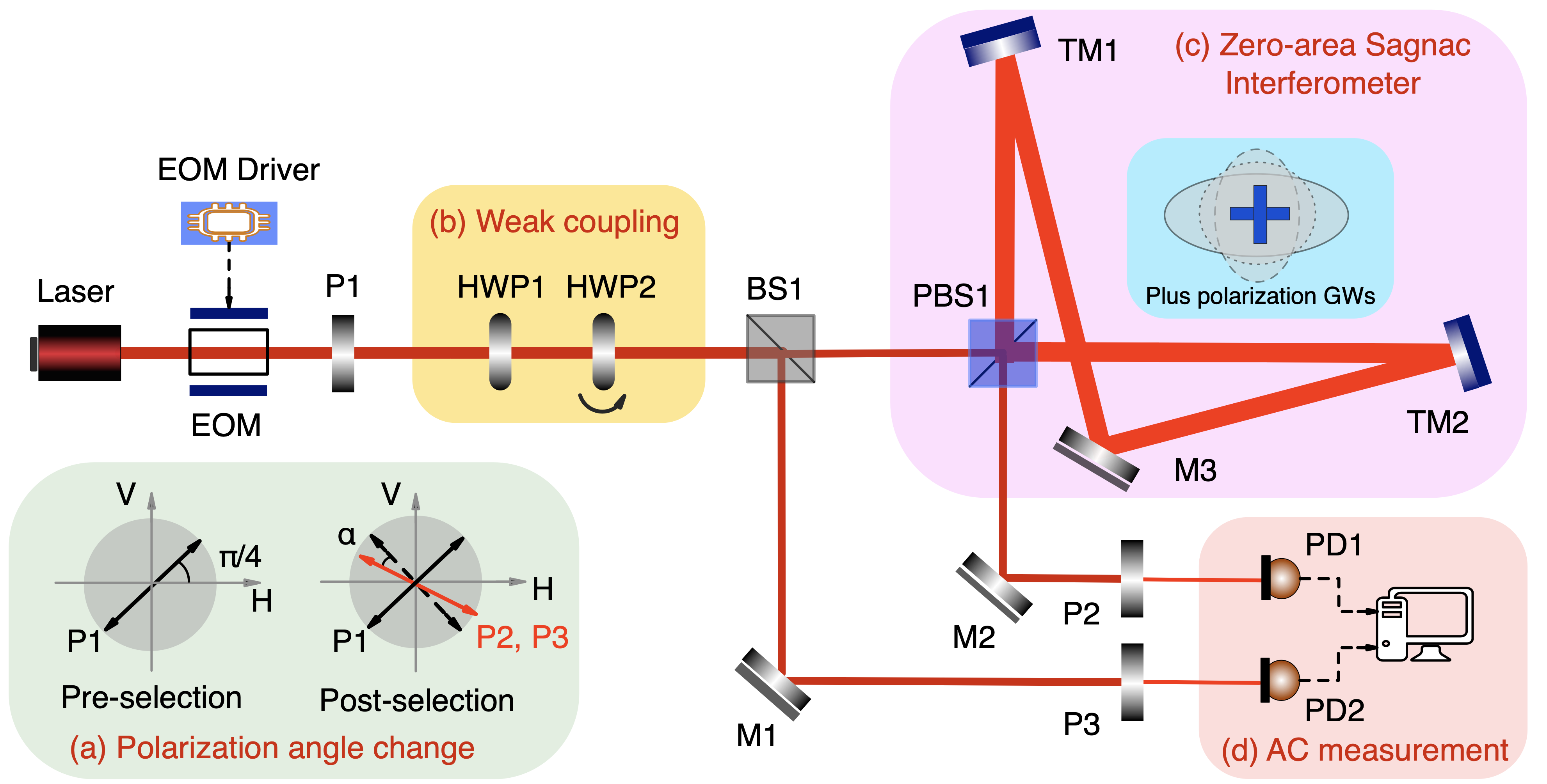}
}
\caption{
The schematic of AWVA for 
the GW detection in the zero-area Sagnac interferometer. 
Laser: a 1064-nm Nd:YAG laser;
EOM: an electro-optic Modulator;
EOM Driver: a high-speed electro-optic modulator driver.
P1, P2 and P3: polarizers;
HWP1 and HWP2: half-wave plates;
BS1: a beam splitter;
PBS1: a polarizing beam splitter;
TM1 and TM2: test masses;
M1, M2 and M3: mirrors;
PD1 and PD2: photodetectors.
(a) The polarization angle of the initial and final state of the system.
(b) The process of ``weak coupling'' in AWVA.
(c) The schematic of the zero-area Sagnac interferometer. 
(d) The process of ``read out'' in AWVA and the estimation of the autocorrelation coefficient.
}
\label{fig:schemeAll}       
\end{figure*}

The zero-area Sagnac interferometer has been experimentally and theoretically investigated for GW detection~\cite{PhysRevLett.104.251102,RevModPhys.86.121,PhysRevLett.76.3053,Grf_2014,Shaddock:98}. 
Figure~\ref{fig:schemeAll}(c) \red{illustrates the configuration where} a ``plus'' oriented GW incident onto the detection plane, \red{with the base tensor components of the GW aligned along the two arms}. Therefore, the absolute value of phase difference $\Delta \phi$ between the two counterpropagating waves can be calculated based on the weak-field approximation~\cite{PhysRevLett.76.3053} of general relativity:
\begin{small}  \begin{eqnarray}
\label{Eq:gravity_phase_difference}
\Delta \phi = 4 \pi f_{l} \left| \int_{t-2\tau_s}^{t-\tau_s} 
\frac{h_{g}e^{i2 \pi f_{g}t}}{2}dt
- \int_{t-\tau_s}^{t} 
\frac{h_{g}e^{i2 \pi f_{g}t}}{2}dt \right | 
= 4f_{l}h_{g}f_{g}^{-1}{\rm sin}^{2}(\pi \tau_{s} f_g) \, , 
\end{eqnarray} \end{small} 
where $f_{l}={c}/{\lambda}$ represents the laser frequency with the wavelength $\lambda$ and speed of light $c$.
\red{The storage time $\tau_s$ is defined as $\tau_s = {2NL}/{c}$, where $L$ represents the arm length and $N$ corresponds to the number of bounces on the end mirror of the delay line. The parameter $\tau_s$ plays a crucial role in determining the location of the peak response within the GW frequency band of interest.
Note that detecting a GW with a strain of $h_g =$ $4 \times 10^{-25}$/$\rm \sqrt{Hz}$ presents both opportunities and challenges, especially considering that the strain noise limits of second-generation interferometers have not yet surpassed $10^{-24}$/$\rm \sqrt{Hz}$~\cite{PhysRevLett.104.251102,RevModPhys.86.121,PhysRevLett.76.3053}}.
Therefore, we proposed a scheme of the GW detection with strain \red{$h_g =$ $4 \times 10^{-25}$/$\rm \sqrt{Hz}$} and simulated the measurements under Gaussian noises with negative SNRs. 

\begin{figure*}[htbp]
	\centering
\subfigure
{
	\begin{minipage}{0.45\linewidth}
\centerline{\includegraphics[scale=0.14,angle=0]{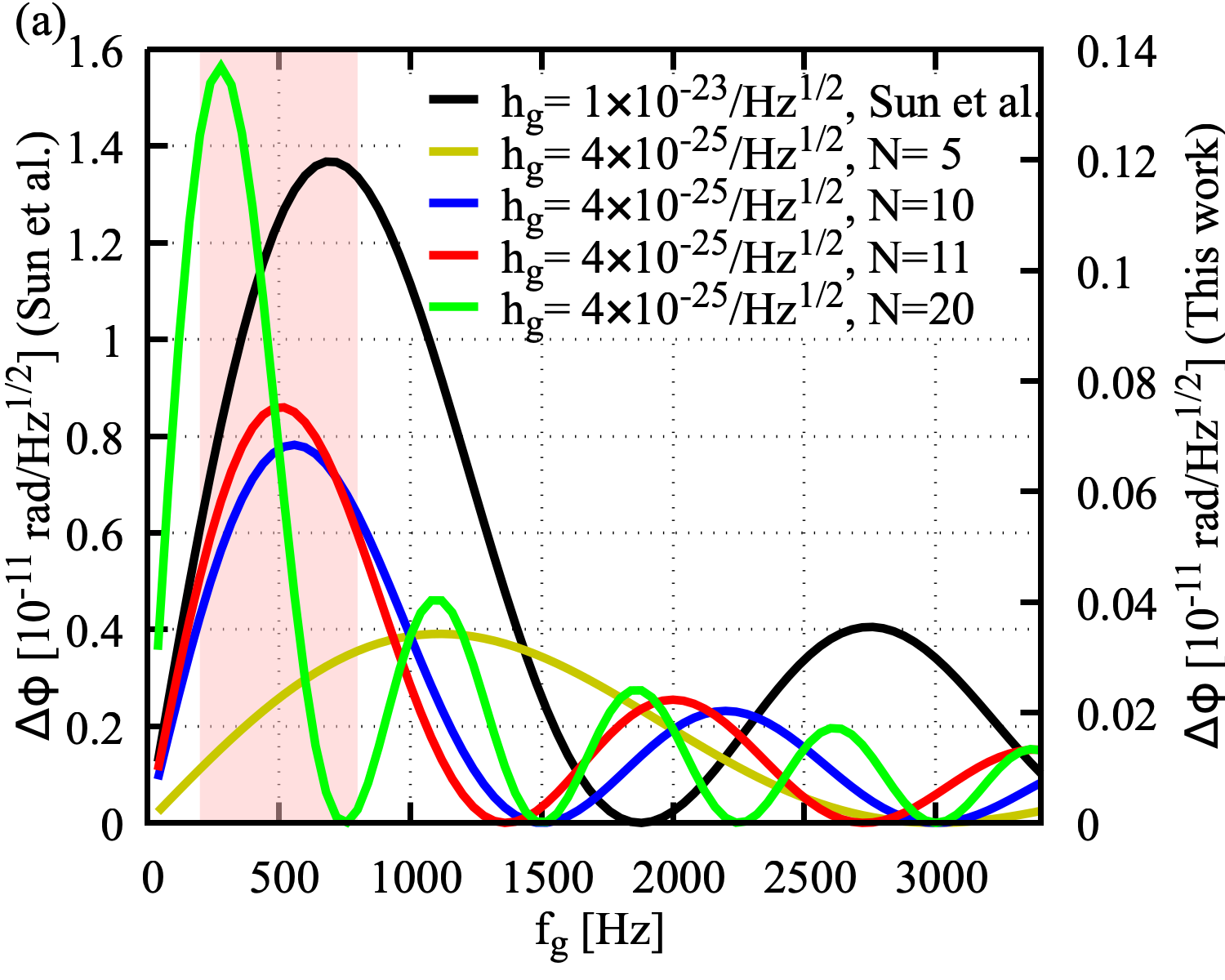}}
	\end{minipage}
}
\subfigure
{
	\begin{minipage}{0.47\linewidth}
\centerline{\includegraphics[scale=0.14,angle=0]{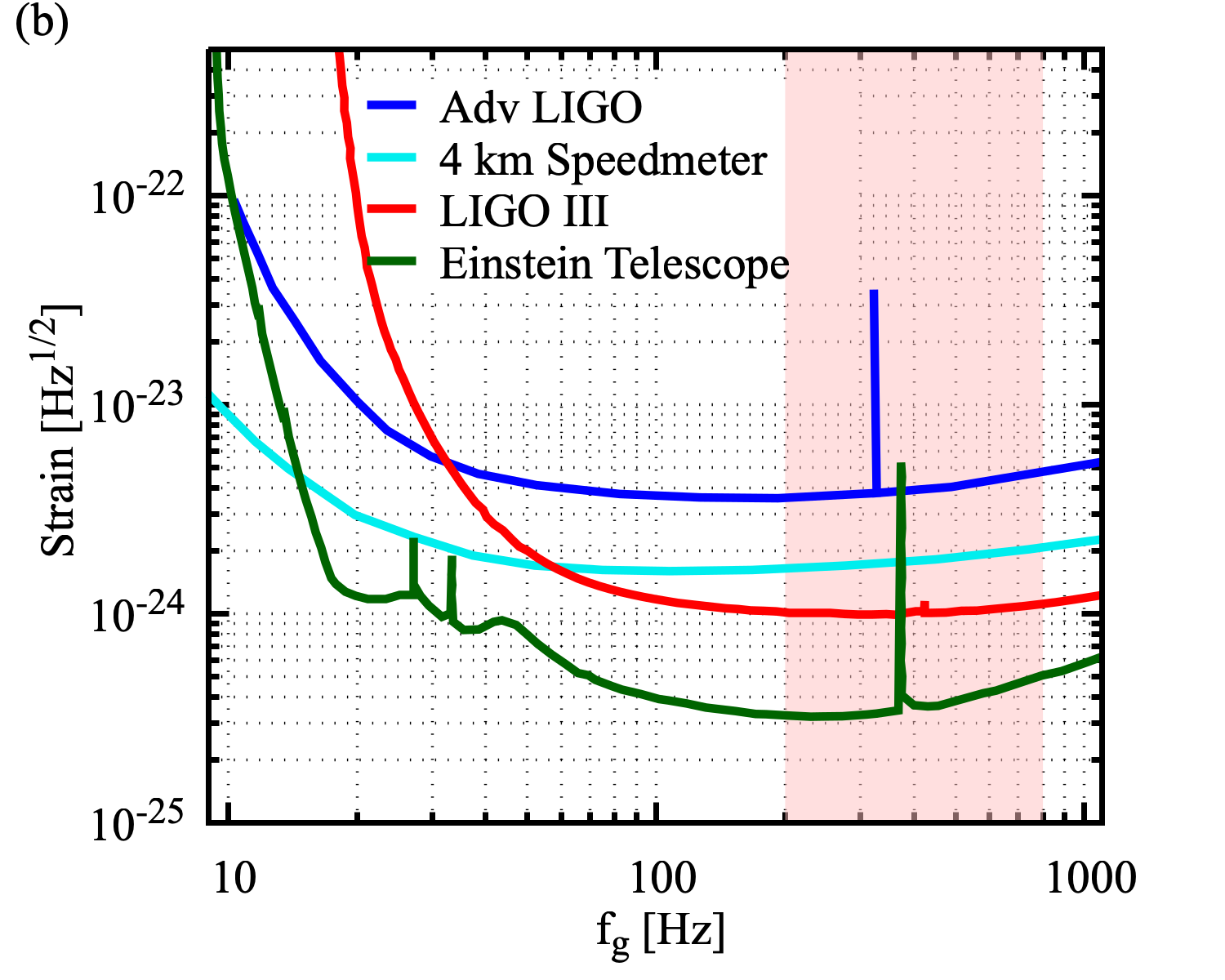}}
	\end{minipage}
}
\caption{\label{Fig:SagnacPhaseAndNoise} 
(a) Theoretical calculation of the phase difference $\Delta \phi$.
\red{
The results with the left vertical coordinate show the calculation of $\Delta \phi$ generated by a GW with strain $h_g =$ $1 \times 10^{-23}$/$\rm \sqrt{Hz}$, in a 20-bounce delay line, 4-km arm length, zero-area Sagnac interferometer with a 1064-nm laser~\cite{PhysRevLett.76.3053}.
The results with the right vertical coordinate show the calculation of $\Delta \phi$ generated by a GW with strain $h_g =$ $4 \times 10^{-25}$/$\rm \sqrt{Hz}$, in different bounce delay line, 10-km arm length, zero-area Sagnac interferometer with a 1064-nm laser.
}
(b) Strain  noise levels of various quantum-enhanced schemes in potential third-generation detectors. Data are taken from Ref.~\cite{RevModPhys.86.121}. 
The pink band of \red{200 Hz $\leq$ $f_g$ $\leq$ 800} \redRR{Hz}  represents the GW band of interest in this paper.
}
\end{figure*}

\red{
The interferometer size is chosen at length $L$ = 10 km to be consistent with the current projects for third-generation GW detectors~\cite{Hall_2019,luck2020design,Maggiore_2020}.
The target peak response $f_{g,max}$ of the third-generation GW detectors is expected to be around $f_{g}$ = 500 Hz.
Note that Eq.~(\ref{Eq:gravity_phase_difference}) indicates that the peak response of the phase difference $\Delta \phi$ depends on the number of bounce $N$ when length $L$ is identified. 
Therefore, we set $N$ = 11 to achieve the desired peak response at $f_{g,max} \approx$ 500 Hz. 
It can be found in Fig.~\ref{Fig:SagnacPhaseAndNoise}(a) that as the number of bounces $N$ increases, the peak response $f_{g,max}$ decreases while the peak value of $\Delta \phi$ increases.
Note that Sun $et \, al.$ also theoretically calculated  $\Delta \phi$ based on Eq.~(\ref{Eq:gravity_phase_difference}) generated by a GW with strain $h_g =$ $1 \times 10^{-23}$/$\rm \sqrt{Hz}$, in a 20-bounce delay line, 4-km arm length, zero-area Sagnac interferometer with a 1064-nm laser~\cite{PhysRevLett.76.3053}.
The results depicted in Fig.~\ref{Fig:SagnacPhaseAndNoise}(a) for different $h_g $ emphasize that smaller $h_g$ values, characteristic of third-generation GW detectors, pose greater challenges in detecting the phase difference $\Delta \phi$ due to its reduced magnitude.
}

\red{The sensitivity and limitations of measuring the phase difference $\Delta \phi$ and the relative strain $h_g$ can be quantitatively described by the strain noise. Figure~\ref{Fig:SagnacPhaseAndNoise}(b) presents the strain noise levels of various quantum nondemolition schemes that are being considered for potential third-generation detectors.}
In this paper, we assume that the amplitude of noise in our simulation remains fixed within the frequency band of \red{200 Hz $\leq$ $f_g$ $\leq$ 800 \redRR{Hz}, as the strain slowly changes over this range}.
Furthermore, numerous researchers have described the noises as the idealized case of stationary-Gaussian noises when analyzing the GW data~\cite{Abbott2020-2,Abbott_2020,Jaranowski2012}.
The simulation results of detecting GW under Gaussian noises with negative-decibel SNRs based on AWVA are shown in Sec.~\ref{Sec:Numericalresult}.

\section{AWVA measurement for the GW detection}
Fig.~\ref{fig:schemeAll} displays the schematic of the autocorrelative weak-value amplification technique employed for GW detection within \red{a} zero-area Sagnac interferometer. The AWVA measurement \red{involves} two simultaneous autocorrelative weak measurements \red{performed on the same initial pointer} in the time domain. \red{These} two weak measurements are conducted by the first beam splitter (BS1) \red{that divides} the light into two paths, as depicted in Fig.~\ref{fig:schemeAll}.
\red{In the upper weak measurement,}
the phase difference $\Delta \phi$ is amplified as one parameter of ``post-selection'' within the zero-area Sagnac interferometer. Meanwhile, another  weak measurement without \red{the phase shift} $\Delta \phi$ is conducted \red{to estimate} the AWVA autocorrelation coefficient. \red{We now provide further details on the application of AWVA for GW detection.}

First, a high-power 1064-nm laser is employed to generate the incident light. Note that ultra-stable high-power laser systems are crucial to achieving a high SNR in various  GW detectors~\cite{RevModPhys.86.121,https://doi.org/10.1002/lpor.200810016,Kwee:12,Bode:20}. \red{For example}, Bode $et \, al.$ demonstrated the robust low-noise operation of the stabilized laser system at 195 W for GW detectors based on solid-state amplifiers~\cite{Bode:20}.
Meijer $et \, al.$ \cite{Meijer:17} reported on a laser system capable of generating temporally shaped pulses at 1064 nm with 0.6 GW of peak power and 44 W of average power by modulation of a continuous-wave seed laser.
Therefore, \red{the initial state $\ket{\Psi_{i}}$ of the pointer} and the state $\ket{\Phi_{i}}$ of the system are prepared by an electro-optic Modulator (EOM) and the first polarizer (P1), respectively:
\begin{equation}
\label{Eq:pre-sel-sys}
\ket{\Phi^{i}_{1,2}}= {\rm sin} \left( \beta \right) \ket{H}+ {\rm cos} \left(\beta \right)\ket{V} \, ,
\end{equation}
\begin{small}  \begin{eqnarray}
\label{Eq:meter_initial_sliding}
|\bkt{t}{\Psi^{i}_{1,2}}|^2(t)=I^{i}_{1,2}
&=&
\frac{1}{\sqrt{2 \pi \sigma_{c} ^{2}}}
e^{-(t+t_{0})^{2}/2\sigma_{c} ^{2}}=I_{0}e^{-(t+t_{0})^{2}/2\sigma_{c} ^{2}} \, ,
\end{eqnarray} \end{small} 
where $\ket{H}$ and $\ket{V}$ represent the horizontal and vertical polarized states. The pre-selected angle is set at $\beta=\pi/4$ as shown in Fig.~\ref{fig:schemeAll}(a).
Parameters $\sigma_{c}$ and $t_0$ \red{represent} the length and the center of the Gaussian pulse, respectively.  The value of $I_0$ denotes the maximum intensity of the \red{temporal pulse}. The subscripts ``1'' and ``2'' represent the upper weak measurement with PD1 and the lower weak measurement with PD2.

Then, the joint state $\ket{\Phi^{i}_{1,2}} \ot \ket{\Psi^{i}_{1,2}}$ of the initial system and pointer evaluates by the time evolution operator: $\hat{U}=e^{-ig\hat{A} \ot \hat{p}} $. Where $\ot $ denotes the tensor product and $g$ represents the strength of weak coupling. The observable operator $\hat{A}$ in this paper is defined as $\hat{A}=\ket{H} \bra{H}-\ket{V} \bra{V} $. The weak coupling strength $g$ is controlled \red{through} the combination of two half-wave plates (HWP1 and HWP2)~\cite{PhysRevLett.111.033604,PhysRevA.103.032212,Fang_2016}:
\begin{equation}
\label{Eq:tinytilt-weakCoupling}
g =  \frac{\lambda_{0}}{2} \left( \frac{1}{\sqrt{1-{\rm sin}^{2} {\eta} /n_{0}^{2} }} -1  \right ) \, ,
\end{equation}
where $\lambda_{0}$ = 1064 nm is the wavelength of the laser and $n_{0}=1.54$ is the refractive index of the HWPs. So the amplitude of $g$ is controlled by the angle $\eta$, which corresponds to the rotation angle of the plane of HWP2 with respect to HWP1.

As shown in Fig.~\ref{fig:schemeAll}, the post-selections of the upper weak measurement and the lower weak measurement are different. In the upper weak measurement with PD1, both the \red{GW-induced} phase difference $\Delta \phi$ and the polarization angle ($\frac{3}{4} \pi$+$\alpha$) of the second polarizer (P2) \red{contribute to} the final state $\ket{\Phi^{f}_{1}}$. At the same time, the final state $\ket{\Phi^{f}_{2}}$ in the lower weak measurement with PD2 is only related to the polarization angle ($\frac{3}{4} \pi$+$\alpha$) of the third polarizer (P3). Therefore, we post-select the system of the two simultaneous autocorrelative weak measurements at \red{two different} states:
\begin{small}  \begin{eqnarray}
\label{Eq:post-sel-sys}
\ket{\Phi^{f}_{1}} &=& 
{\rm sin} \left( \frac{3}{4 }\pi + \alpha \right) {\rm e} ^{i\Delta \phi  }\ket{H}
               + {\rm cos} \left( \frac{3}{4 }\pi + \alpha \right)  \ket{V}, \\
\ket{\Phi^{f}_{2}} &=& {\rm sin} \left( \frac{3}{4 }\pi + \alpha \right) \ket{H}
               + {\rm cos} \left( \frac{3}{4 }\pi + \alpha \right)  \ket{V}.
\end{eqnarray} \end{small} 
So the weak value of the two weak measurements can be evaluated by:
\begin{small}  \begin{eqnarray}
\label{Eq:weak_value12}
 A_{w1}&=&  \frac{\bra{\Phi^{f}_1}\hat{A}\ket{\Phi^{i}_1}}{\bkt{\Phi^{f}_1}{\Phi^{i}_1}} 
= \frac{[{\rm cos}(\alpha)-{\rm sin}(\alpha)]e^{-i\Delta \phi}-[-{\rm cos}(\alpha)-{\rm sin}(\alpha)]}{[{\rm cos}(\alpha)-{\rm sin}(\alpha)]e^{-i\Delta \phi}+[-{\rm cos}(\alpha)-{\rm sin}(\alpha)]}  \,, \\
A_{w2}&=& \frac{\bra{\Phi^{f}_2}\hat{A}\ket{\Phi^{i}_2}}{\bkt{\Phi^{f}_2}{\Phi^{i}_2}} 
= \frac{[{\rm cos}(\alpha)-{\rm sin}(\alpha)]-[-{\rm cos}(\alpha)-{\rm sin}(\alpha)]}{[{\rm cos}(\alpha)-{\rm sin}(\alpha)]+[-{\rm cos}(\alpha)-{\rm sin}(\alpha)]}  \,. 
\end{eqnarray} \end{small} 
Note that $ A_{w1}$ is a complex number while $ A_{w2}= -{{\rm cos}(\beta)}/{{\rm sin}(\beta)}$ is a pure real number.
According to the principle of weak-value amplification~\cite{PhysRevA.85.052110}, the weak measurement with an arbitrary coupling strength $g$ leads to temporal shifts $\Delta \tau_{1}$ and $\Delta \tau_{2}$ of the pointers in the time domain:
\begin{small}  \begin{eqnarray}
\label{Eq:timeShifts1}
\Delta \tau_{1} &=& \frac{{\rm Re}[A_{w1}]g/c}{1+\frac{1}{2}(1-|A_{w1}|^2)(e^{-2g^{2}/(c^{2}\sigma_{c}^{2})} -1)}  \,,
\end{eqnarray} \end{small} 
\vspace*{-4mm} 
\begin{small}  \begin{eqnarray}
\label{Eq:timeShifts2}
\Delta \tau_{2} &=& \frac{{\rm Re}[A_{w2}]g/c}{1+\frac{1}{2}(1-|A_{w2}|^2)(e^{-2g^{2}/(c^{2}\sigma_{c}^{2})} -1)} \,.
\end{eqnarray} \end{small} 
Meanwhile, \red{the post-selection of WVA} also leads to the attenuations $\Pi_{w1}=|\bkt{\Phi^{f}_1}{\Phi^{i}_1}|^{2}=[1-{\rm cos}(2\alpha){\rm cos}(\Delta \phi)]/2$ and $\Pi_{w2}=|\bkt{\Phi^{f}_2}{\Phi^{i}_2}|^{2}= {\rm sin}^2(\alpha) $ in intensity. Considering the fact of $\Delta \phi$ $\ll$ $\alpha$, we take the approximation \red{$\Pi_{w1} \approx \Pi_{w2}= {\rm sin}^2(\alpha)$} in this paper.
Note that the attenuations $\Pi_{w1}$ and $\Pi_{w2}$ are also called the probability of the successful post-selection in AWVA.
With the given pre- and post-selected states of the system in the two weak measurements, we can obtain the final signals $I^{f} _{1}(t)$ and $I^{f} _{2}(t)$ detected on PD1 and PD2 respectively:
\begin{small}  \begin{eqnarray}
\label{Eq:meter_final_pointer1}
I^{f} _{1}(t)&=& \Pi_{w1} I^{i} _{1}(t-\Delta \tau_{1})
={\rm sin}^2(\alpha) I_{0} 
e^{-(t+t_{0}-\Delta \tau_{1})^{2}/2\sigma_{c} ^{2}}, 
\end{eqnarray} \end{small} 
\vspace*{-4mm} 
\begin{small}  \begin{eqnarray}
\label{Eq:meter_final_pointer2}
I^{f} _{2}(t) &=& \Pi_{w2} I^{i} _{2}(t-\Delta \tau_{2})= {\rm sin}^2(\alpha) I_{0} 
e^{-(t+t_{0}- \Delta \tau_{2})^{2}/2\sigma_{c} ^{2}}.
\end{eqnarray} \end{small} 
Particularly, Gaussian white noises ${N}(t,\xi_{1})$ and ${N}(t,\xi_{2})$ with different seeds (corresponding to $\xi_{1}$ and $\xi_{2}$) are added to the final signals $I^{f} _{1}(t)$ and $I^{f} _{2}(t)$ to simulate the GW detection in the \redRR{ideal} case with only Gauussin noises souces~\cite{Abbott2020-2,Abbott_2020,Jaranowski2012}. Note that different seeds also indicate that we only consider the uncorrelated noises in the AWVA scheme. In this case, the autocorrelation coefficient $\Theta_{N}$(t) under noise can be calculated \red{by applying signal denoising techniques commonly used in engineering}~\cite{lee1949application,machines9060123}:
\begin{small}  \begin{eqnarray}
\label{Eq:ACIdefine+noise}
{\rm \Theta}_{N}(t;f_g) &=& \int_{0}^{t}  \left [ I_{1}^{f}(t^{\prime}; \Delta \tau_{1})+{N}(t^{\prime},\xi_{1}) \right ] \times  \left [ I_{2}^{f}(t^{\prime}, \Delta \tau_{2})+{N}(t^{\prime},\xi_{2})\right ] dt^{\prime}  \,  \nonumber\\
&\approx&\int_{0}^{t}  I_{1}^{f}(t^{\prime}; \Delta \tau_{1}) \times  I_{2}^{f}(t^{\prime}; \Delta \tau_{2})  dt^{\prime}\, ,
\end{eqnarray} \end{small} 
where the \red{approximation} holds when the correlation ${\rm \Theta}_{IN}(t)$ and ${\rm \Theta}_{NI}(t)$ between signal and noise, and the correlation ${\rm \Theta}_{NN}(t)$ between noises \red{satisfy the conditions}: ${\rm \Theta}_{IN}(t)=\int_{0}^{t}  I_{1}^{f}(t^{\prime};\Delta \tau_{1}) \times {N}(t^{\prime},\xi_{2}) dt^{\prime} \approx 0$, ${\rm \Theta}_{NI}(t)=\int_{0}^{t}  {N}(t^{\prime},\xi_{1}) \times  I_{2}^{f}(t^{\prime}; \Delta \tau_{2}) \approx 0\,$ and ${\rm \Theta}_{NN}(t)=\int_{0}^{t}  {N}(t^{\prime},\xi_{1}) \times {N}(t^{\prime},\xi_{2})\approx 0\,$. It can be found  in our previous work~\cite{AWVA} that the mean value of the autocorrelation coefficient $\Theta_{N}$(t) with multiple measurements \red{is consistent with} the theoretical value $\Theta_{0}$(t) without noises. 
Note that the value of ${\rm \Theta}_{N}(t)$ depends on the integration time $t$. So the choice of \red{the initial integration time will significantly impact the outcome.}
In addition, both the temporal shift $\Delta \tau_{2}$ and $ I_{2}^{f}(t^{\prime}; \Delta \tau_{2})$ remain unchanged by the influence of GW events.
Therefore, by using signal $ I_{2}^{f}(t^{\prime}; \Delta \tau_{2})$ to trigger acquiring signal and \red{perform the integration},  the calculation ${\rm \Theta}_{N}(t)$ of Eq.~(\ref{Eq:ACIdefine+noise}) is converted into an integral:
\begin{small}  \begin{eqnarray}
\label{Eq:ACIdefine+noiseNormolized}
{\rm \Theta}_{N}(t;f_g) 
&=&\int_{0}^{t}  I_{1}^{f}(t^{\prime}; \Delta \tau_{1}-\Delta \tau_{2}) \times  I_{2}^{f}(t^{\prime})  dt^{\prime} \\ 
&=&
\red{\int_{0}^{t} \left[  e^{-(t^{\prime}+t_{0}-(\Delta \tau_{1}- \Delta \tau_{2}))^{2}/2\sigma_{c} ^{2}} \right ] \times \left[ 
e^{-(t^{\prime}+t_{0})^{2}/2\sigma_{c} ^{2}} \right ] dt^{\prime}\,.} \nonumber
\end{eqnarray} \end{small} 
Eq.~(\ref{Eq:ACIdefine+noiseNormolized}) indicates that the light intensity $I_{1}^{f}(t^{\prime}; \Delta \tau_{1}-\Delta \tau_{2})$ and $I_{2}^{f}(t^{\prime}) $ are normalized by \red{setting ${\rm sin}^2(\alpha) I_{0}$= 1} in this paper.
Based on the investigation of AWVA~\cite{AWVA}, the sensitivity of measuring ${\rm \Theta}_{N}(t)$ depends on the integration time t.
The sensitivity $\kappa$ is difined as $\kappa(f_g)$= $\mid{\rm \Theta}_{N}(t;f_g)$ - ${\rm \Theta}_{0}(t;f_g=0)\mid$/$f_g$. The sensitivity $\kappa$ is maximized when the integration time $t$ is set to $t_0$.
\red{Where ${\rm \Theta}_{0}(t;f_g)$ is defined and calculated as the result of the measurement without noises.}
The simulation results of ${\rm \Theta}_{N}(t;f_g) $ and $\kappa(f_g)$ are displayed in Sce.~\ref{Sec:Numericalresult}.

\section{Numerical result and discussion}
\label{Sec:Numericalresult}
In the following, we show the simulation results of AWVA for the GW detection with GW strain of \red{$h_g =$ $4 \times 10^{-25}$/$\rm \sqrt{Hz}$} under Gaussian white noises with negative SNRs. Before the simulation, there are several parameters in AWVA need to be determined:
\begin{itemize}
\vspace{-0.0cm}
\item[(i)] 
The size of the zero-area Sagnac interferometer has been discussed in Sec.~\ref{sec:zero-areaSagnac} and is taken from \red{the design of the third-generation Einstein Telescope GW detector~\cite{Maggiore_2020}, with an arm length $L$ of 10 km and wavelength of light $\lambda$ of 1064 nm. In order to obtain the peak response at $f_{g, max}$ = 500 Hz, the number of bounces is set at $N$ = 11. The dependency of the GW-induced phase difference $\Delta \phi$ on the GW frequency $f_g$ is shown in Fig.~\ref{Fig:SagnacPhaseAndNoise}(a). }
\begin{figure*}[b]
\centering
 \subfigure
{
	\begin{minipage}{0.45\linewidth}
\centerline{\includegraphics[scale=0.14,angle=0]{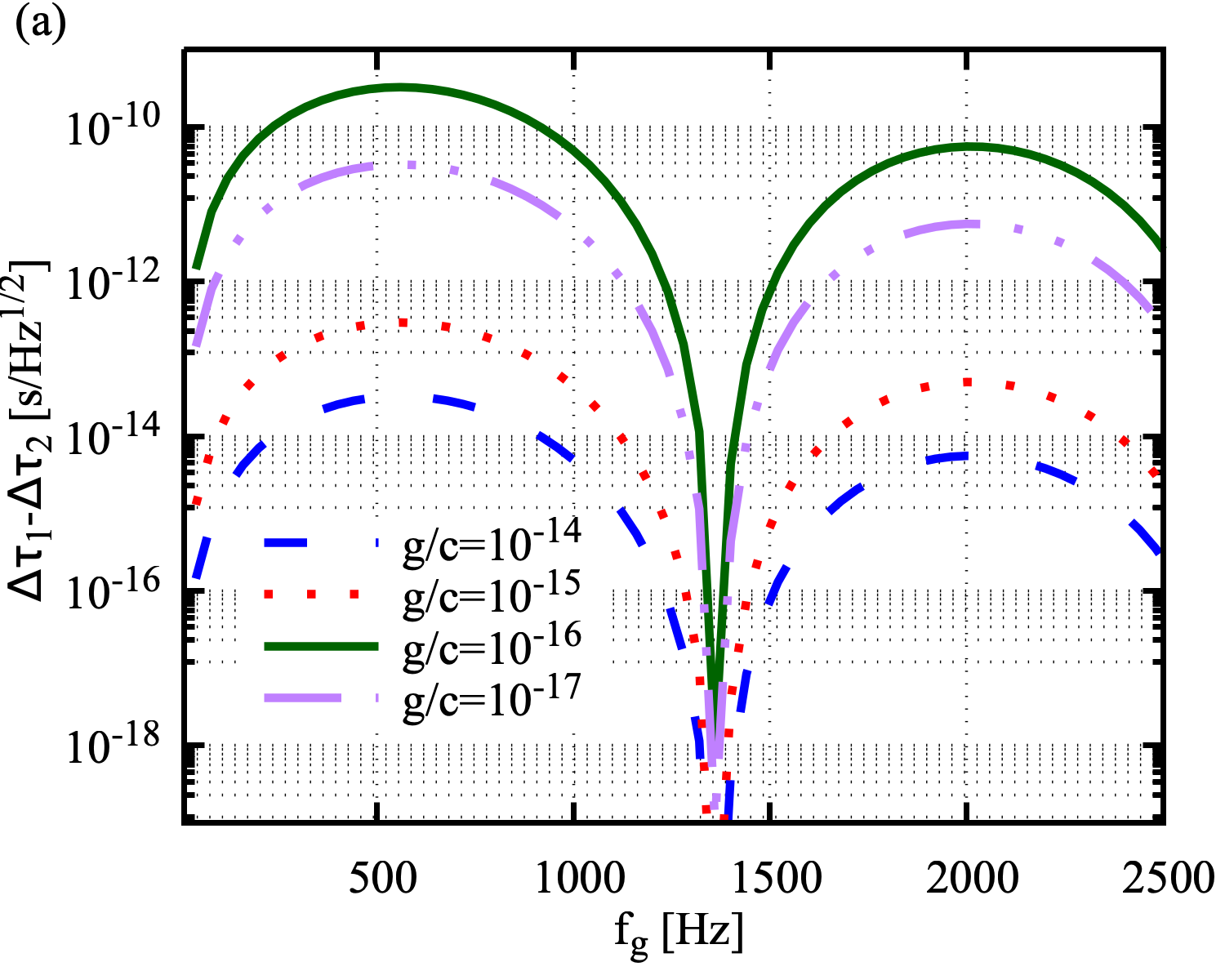}}
	\end{minipage}
}
\subfigure
{
	\begin{minipage}{0.45\linewidth}
\centerline{\includegraphics[scale=0.14,angle=0]{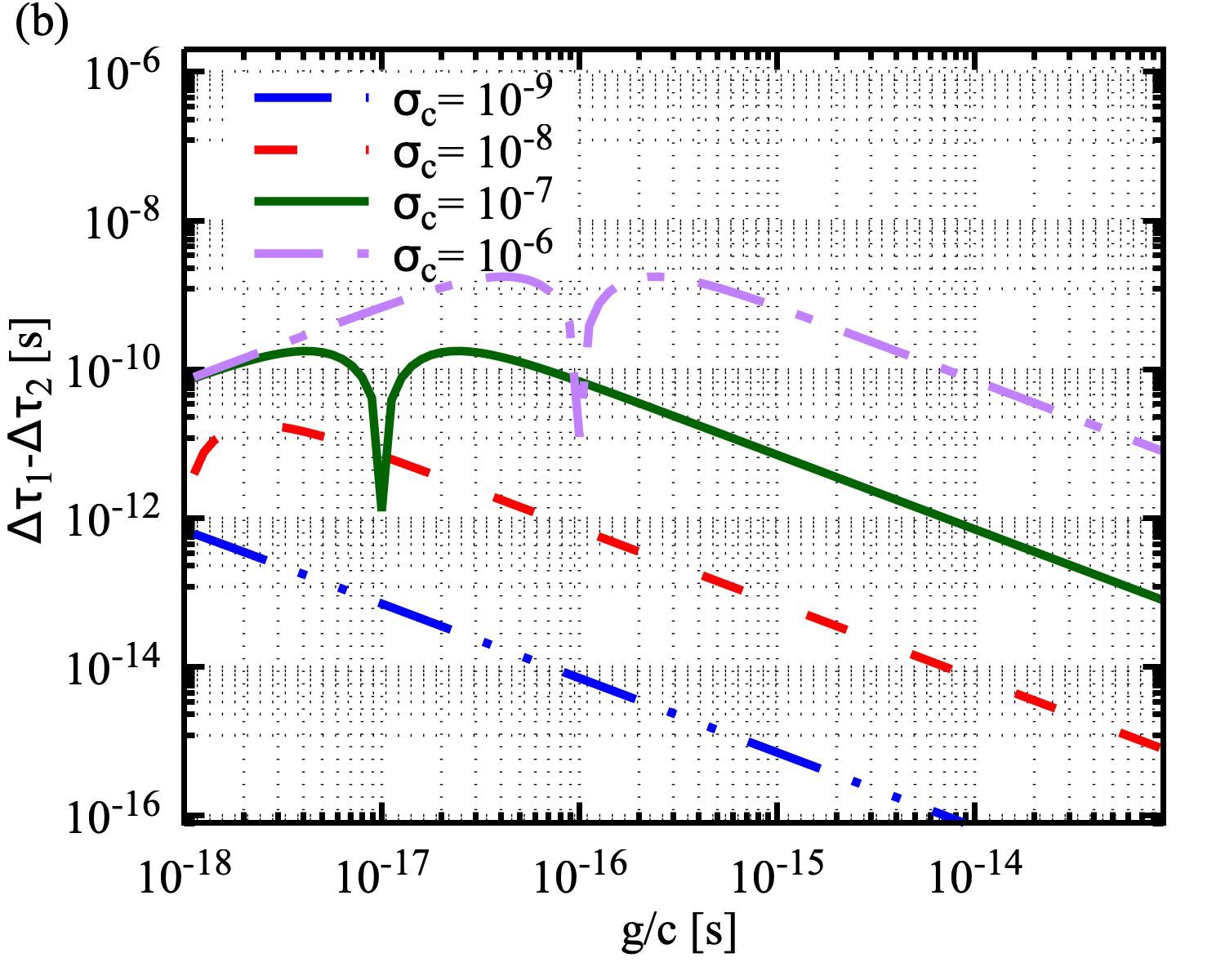}}
	\end{minipage}
}
\caption{\label{Fig:ChosingBestParameters} The temporal shifts of $\Delta \tau_{1}-\Delta \tau_{2}$ of measuring a GW with strain \red{$h_g =$ $4 \times 10^{-25}$/$\rm \sqrt{Hz}$}.
(a) The dependence of the temporal shifts of $\Delta \tau_{1}-\Delta \tau_{2}$ on $f_g$  for different $g/c$, with a \red{fixed pulse length of $\sigma_{c}=$ 100 ns}.
(b) The dependence of the temporal shifts $\Delta \tau_{1}-\Delta \tau_{2}$ on $g/c$ for various pulse lengths $\sigma_{c}$ when detecting a GW at the maximum frequency $f_{g, {max}} = 500$ Hz.
}
\end{figure*}
\item[(ii)] 
\red{In AWVA, we determine the value of $\alpha$ based on the amplitude of probability $\Pi_{w2}$ of successful post-selection, rather than adjusting the polarized angle between P2 and P3.}
This approach has also been employed in the weak measurement \red{ to demonstrate} ultra-low noise phase in a fiber-based Sagnac interferometer~\cite{Liu:22}.
Therefore, \red{considering a laser with 1 GW  peak power~\cite{Meijer:17} and PDs with the lowest resolvable power of 10 pW (corresponding to noise-equivalent power~\cite{mackowiak2015nep}), the probability $\Pi_{w2}$ of the successful post-selection will be as low as $10^{-20}$.
Correspondingly, the post-selected angle $\alpha$ can be calculated and determined \red{as} $\alpha=$ $10^{-10}$ rad.
}

\item[(iii)] 
The pulse length $\sigma_{c}$ is set \red{to} $\sigma_{c}$= $10^{-6}$ s, which is obtained by considering the largest temporal shifts $\Delta \tau_{1}-\Delta \tau_{2}$ when detecting a GW at  $f_{g, max}$ = \red{500} Hz. \red{Specifically}, we \red{present the dependence of temporal shifts $\Delta \tau_{1}-\Delta \tau_{2}$ on the GW strain $f_g$ for different AWVA coupling strengths $g/c$ and a fixed pulse length of $\sigma_{c}=100$ ns in} Fig.~\ref{Fig:ChosingBestParameters}(a). 
The change of curves $\Delta \tau_{1}-\Delta \tau_{2}$ dependence of $f_g$ qualitatively corresponds to the change of the phase difference $\Delta \phi$ in Fig.~\ref{Fig:SagnacPhaseAndNoise}(a).
{Similarly,} the peak response of $\Delta \tau_{1}-\Delta \tau_{2}$ also accurs aroud $f_{g, max}$ = \red{500} Hz.
Then, we further calculate \red{the dependence of the temporal shifts $\Delta \tau_{1}-\Delta \tau_{2}$ on $g/c$ and $\sigma_{c}$} when detecting a GW at frequency $f_{g, max}$ = \red{500} Hz in Fig.~\ref{Fig:ChosingBestParameters}(b). 
The results indicate that the larger $\sigma_{c}$ leads to more significant temporal shifts  $\Delta \tau_{1}-\Delta \tau_{2}$. However, the pulse length $\sigma_{c}$ can not be too large because the measurement time $MT_{A}$ \red{of multiple AWVA measurements} must be much smaller than the period (1/$f_g$) of the detected GWs. Where $M$ represents the number of repeated measurements for obtaining smaller error bars of $\Theta$~\cite{AWVA}. In this paper, the center of the Gaussian pulse is set at $t_0$ = 3 \redRR{$\upmu$s}, \red{resulting in} $T_{A}$ = 6 ${\rm \Theta}_{N}(t=3\,$ \redRR{$\upmu$s}) for a single measurement.
%
%
\item[(iv)] The AWVA coupling strength $g/c$ is chosen \red{to be} $g/c$ = $10^{-16.3}$ s when $\sigma_{c}$ = $10^{-6}$ s. 
Based on the nonlinear relations Eqs.~(\ref{Eq:timeShifts1}) and (\ref{Eq:timeShifts2}) as well as Fig.~\ref{Fig:ChosingBestParameters}(b), it is evident that increasing AWVA coupling strength $g/c$ does not always \red{result in an enhancement of the shift}  $\Delta \tau_{1}-\Delta \tau_{2}$. 
The first peak response of $\Delta \tau_{1}-\Delta \tau_{2}$ occurs at $g/c$ = $10^{-16.3}$ s. Note that the value of $\Delta \tau_{1}-\Delta \tau_{2}$ also depends on the GW frequency $f_g$, as shown in Fig.~\ref{Fig:ChosingBestParameters}(a). However, in this paper, we have neglected the influence of GW frequency $f_g$ on the choice of $g/c$.
\end{itemize}

\begin{figure*}[b]
\centering
 \subfigure
{
	\begin{minipage}{0.31\linewidth}
\centerline{\includegraphics[scale=0.14,angle=0]{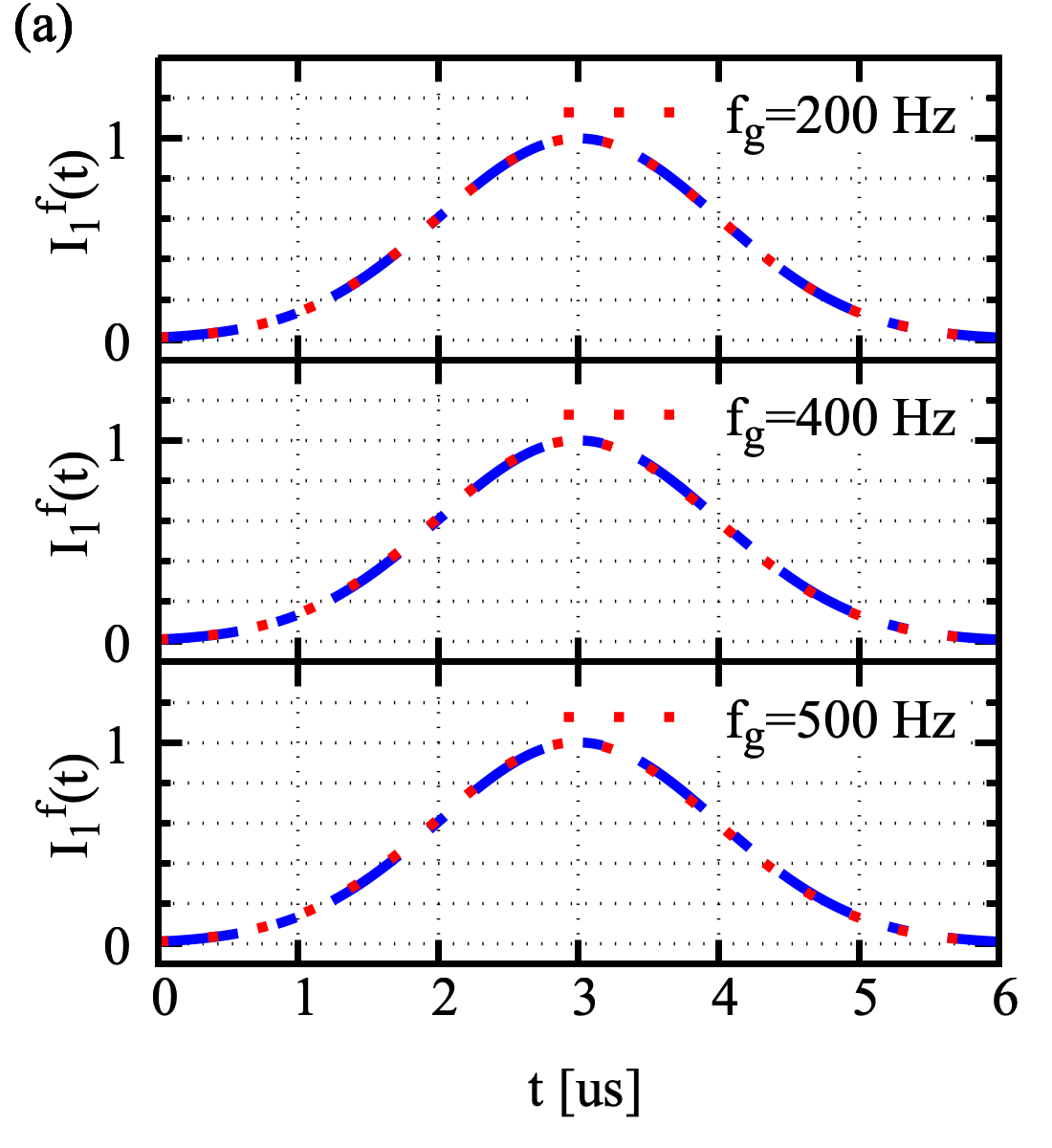}}
	\end{minipage}
}
\subfigure
{
	\begin{minipage}{0.31\linewidth}
\centerline{\includegraphics[scale=0.14,angle=0]{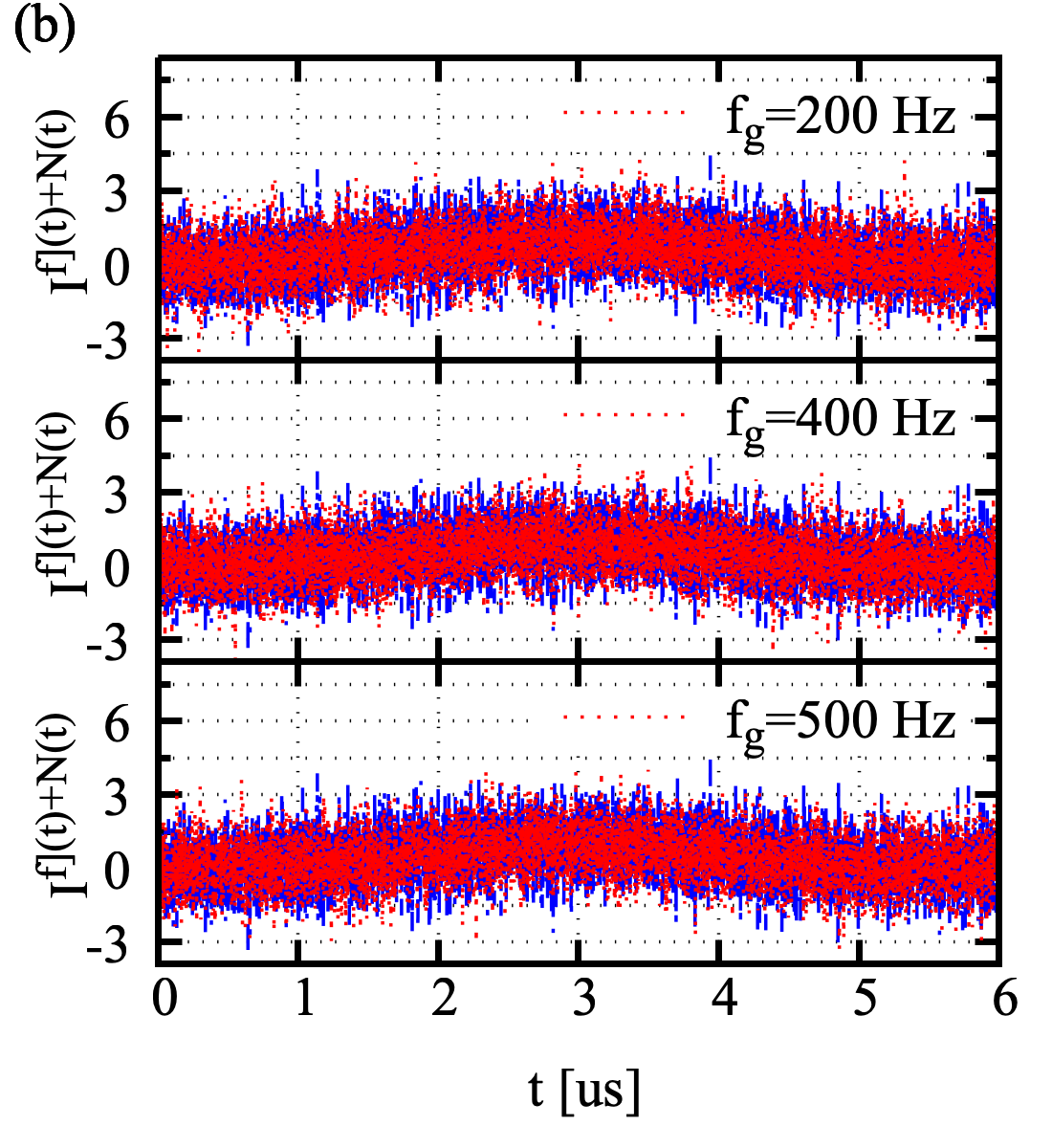}}
	\end{minipage}
}
\subfigure
{
	\begin{minipage}{0.31\linewidth}
\centerline{\includegraphics[scale=0.14,angle=0]{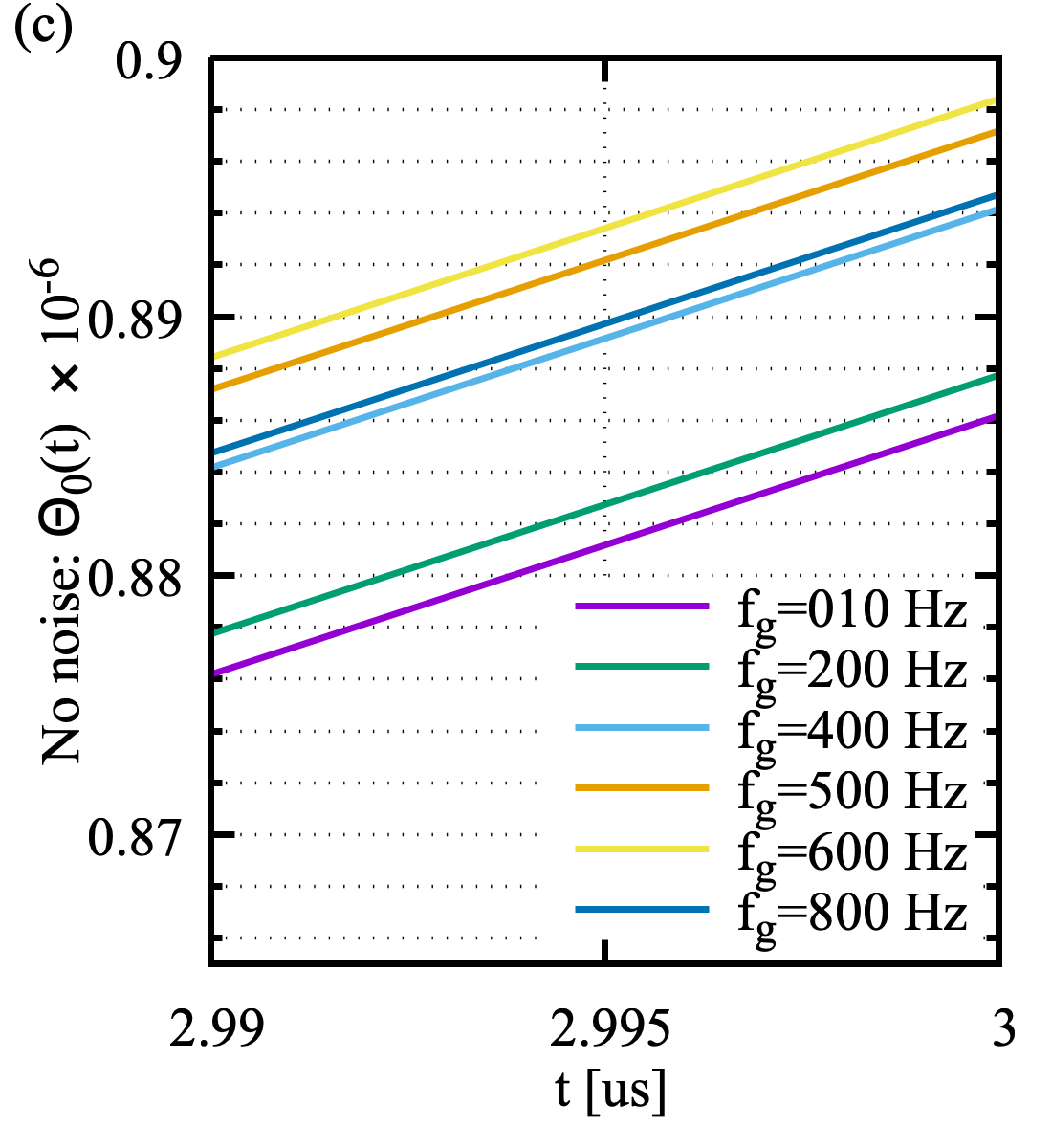}}
	\end{minipage}
}
\caption{\label{Fig:Pointer_shiftCompaaring} 
The final pointer $I^{f}_{1}(t)$ (a) and the mix signals (b) of $I^{f}_{1}(t)$ and Gaussian noise $N(t)$ at SNR = -12 dB.
The data in blue represents the results at $f_g$ = 0 Hz.
\red{ (c) The ideal curves of ${\rm \Theta}_{N}(t)$ for GW detection with $h_g =$ $4 \times 10^{-25}$/$\rm \sqrt{Hz}$ without noises. }
}
\end{figure*}
After setting the above parameters, we proceed to simulate the AWVA measurements for GW detection with GW strain of \red{$h_g =$ $4 \times 10^{-25}$/$\rm \sqrt{Hz}$}. In particular, we perform the simulation of measuring ${\rm \Theta}_{N}(t;f_g) $ \red{at} different GW frequencies $f_g$, \red{using} \red{three} sampling frequencies $f_D$ of PDs, and \red{considering} noises with different negative-decibel SNRs. Following the definition of SNR in Ref.~\cite{AWVA}, we have ${\rm SNR}=20 \, {\rm log}_{10} {{\rm Max}[I^{f}_{1,2}(t)] }/{{\rm Max}[N(t)]} $. Where ${{\rm Max}[I^{f}_{1,2}(t)] }$ and ${{\rm Max}[N(t)]}$ represent the maximum values of the signal and the noise, respectively. 

\begin{figure*}[t]
\centering
 \subfigure
{
	\begin{minipage}{0.45\linewidth}
\centerline{\includegraphics[scale=0.14,angle=0]{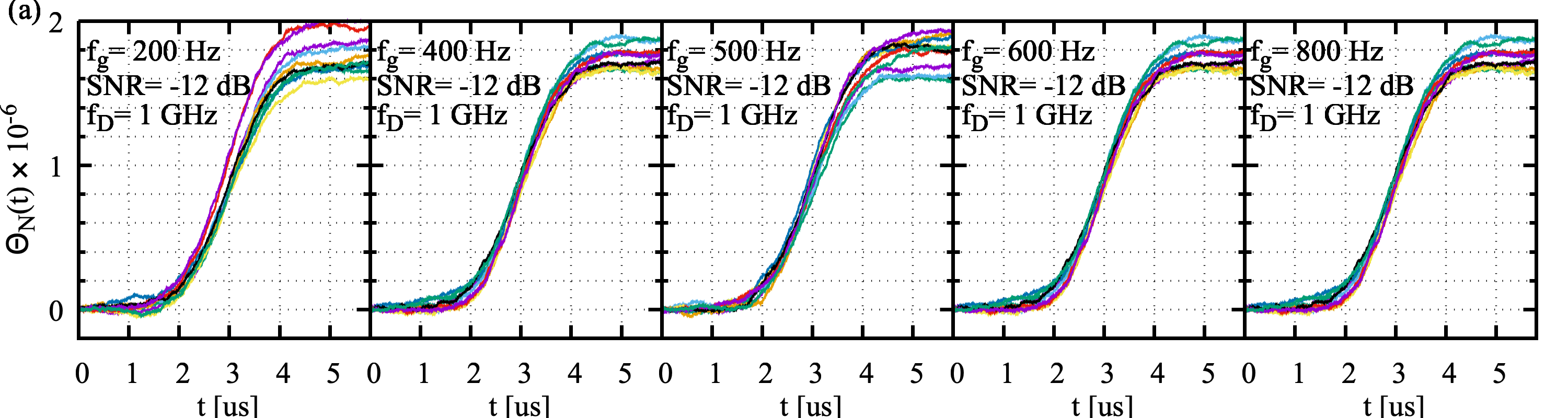}}
	\end{minipage}
}

\subfigure
{
	\begin{minipage}{0.45\linewidth}
\centerline{\includegraphics[scale=0.14,angle=0]{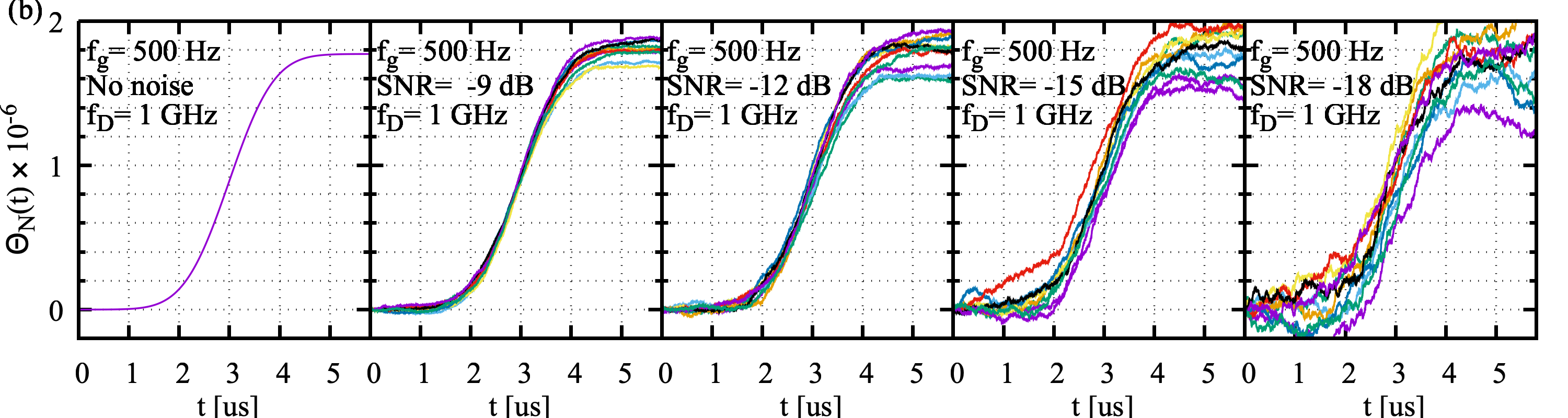}}
	\end{minipage}
}

\subfigure
{
	\begin{minipage}{0.45\linewidth}
\centerline{\includegraphics[scale=0.14,angle=0]{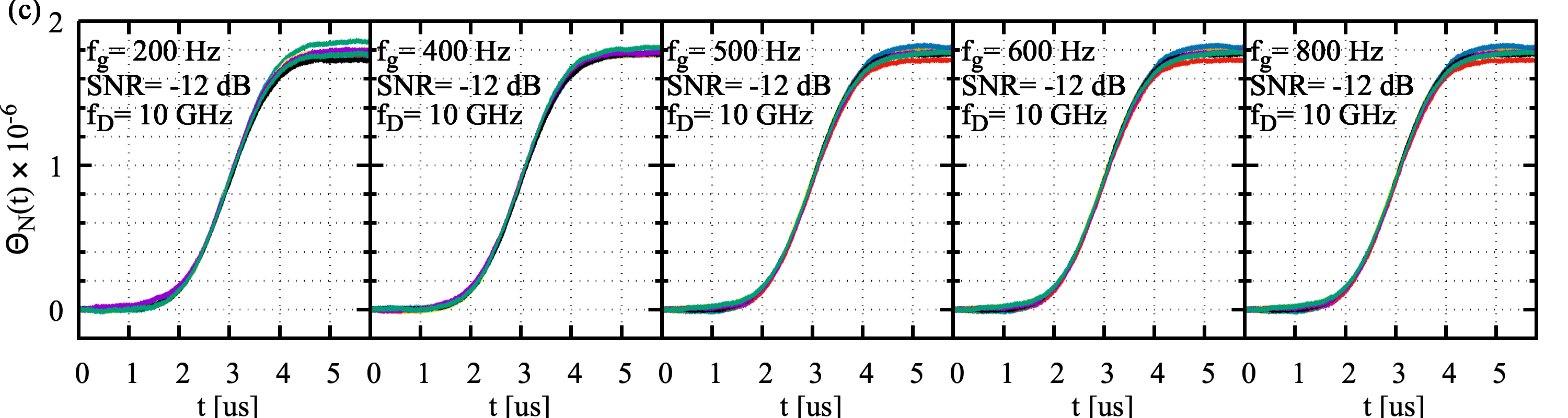}}
	\end{minipage}
}

\subfigure
{
	\begin{minipage}{0.45\linewidth}
\centerline{\includegraphics[scale=0.14,angle=0]{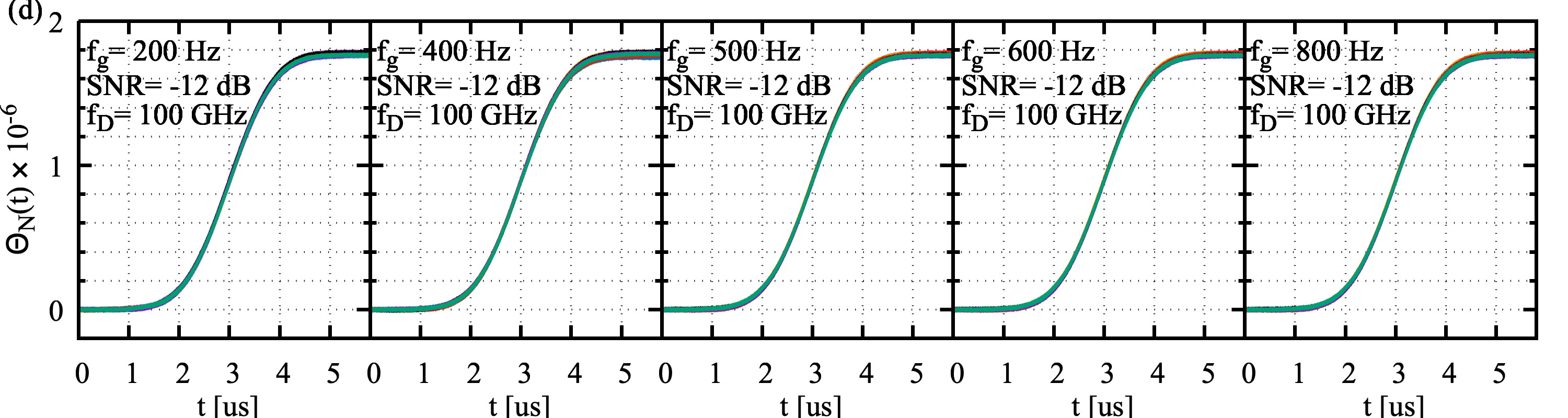}}
	\end{minipage}
}
\caption{\label{Fig:AWVA_coffience_shiftCompaaring}
The simulation of ${\rm \Theta}_{N}(t)$ for GW detection with strain \red{$h_g =$ $4 \times 10^{-25}$/$\rm \sqrt{Hz}$}. 
(a) The measurements with different $f_g$ at SNR= -12 dB and $f_D=$ 1 GHz.
(b) The measurements under noises with various SNRs.
(c) The measurements with different $f_g$ at SNR= -12 dB and $f_D=$ 10 GHz.
\red{
(d) The measurements with different $f_g$ at SNR= -12 dB and $f_D=$ 100 GHz.
}
The lines with different colors in each figure represent the measurements with noises at different random seeds $\xi$.
}
\end{figure*}

Figs.~\ref{Fig:Pointer_shiftCompaaring}(a) and \ref{Fig:Pointer_shiftCompaaring}(b) show the final pointer $I^{f}_{1}(t)$ and the mix signals of $I^{f}_{1}(t)$ and Gaussian noise $N(t)$ at SNR = -12 dB. \red{Figs.~\ref{Fig:Pointer_shiftCompaaring} (c) shows the ideal curves of ${\rm \Theta}_{N}(t)$ for GW detection with different $f_g$ without noises.}
Note that directly detecting the pointer shifts \red{in the presence of background noise is challenging}. 
However, AWVA offers the potential to detect temporal shifts in the presence of Gaussian noise, even at negative SNRs, as demonstrated in previous studies~\cite{AWVA}. Therefore, we display the simulation results of ${\rm \Theta}_{N}(t)$ with different setups in Fig.~\ref{Fig:AWVA_coffience_shiftCompaaring}.

\begin{figure*}[t]
\centering
 \subfigure
{
	\begin{minipage}{0.31\linewidth}
\centerline{\includegraphics[scale=0.14,angle=0]{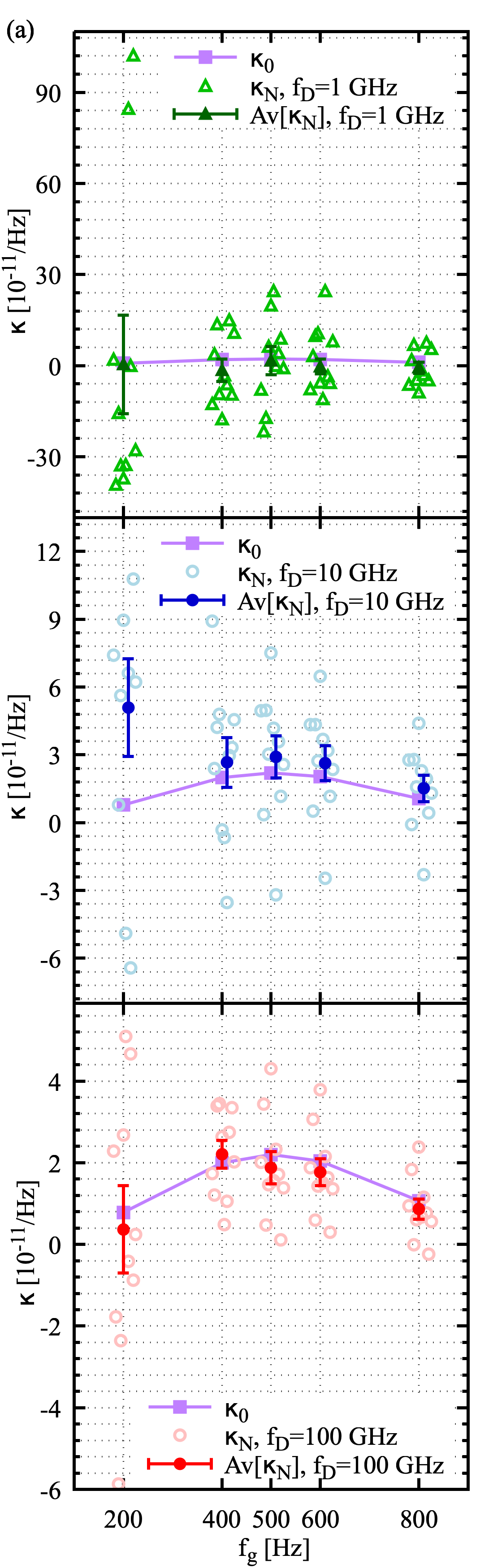}}
	\end{minipage}
}
\subfigure
{
	\begin{minipage}{0.31\linewidth}
\centerline{\includegraphics[scale=0.14,angle=0]{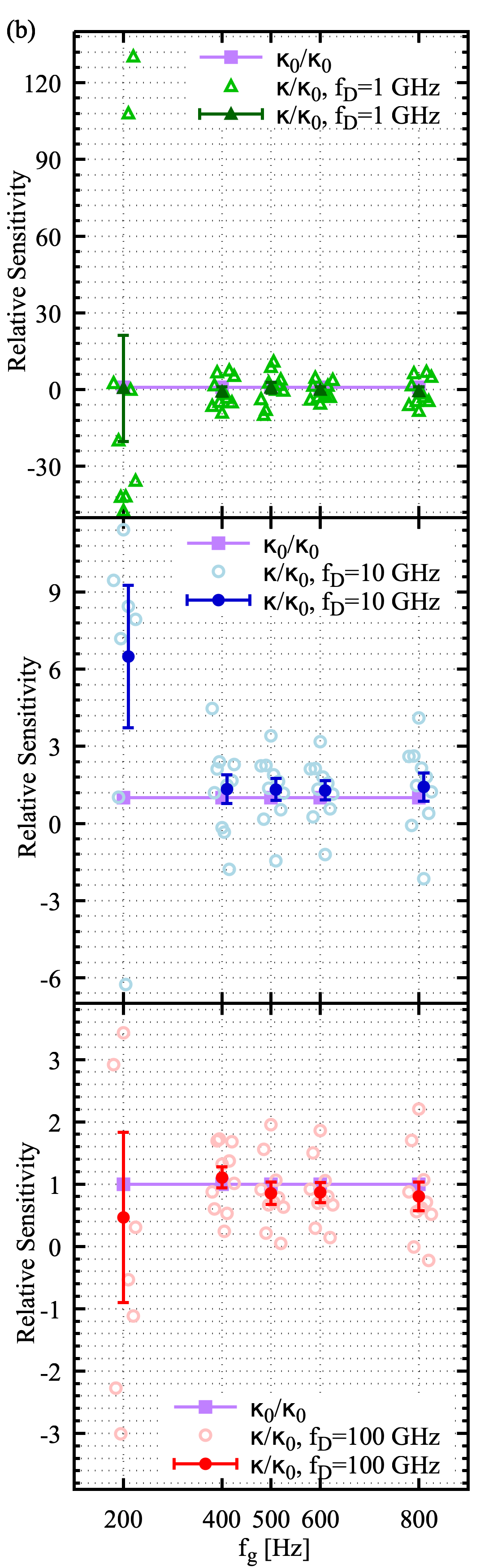}}
	\end{minipage}
}
\subfigure
{
	\begin{minipage}{0.31\linewidth}
\centerline{\includegraphics[scale=0.14,angle=0]{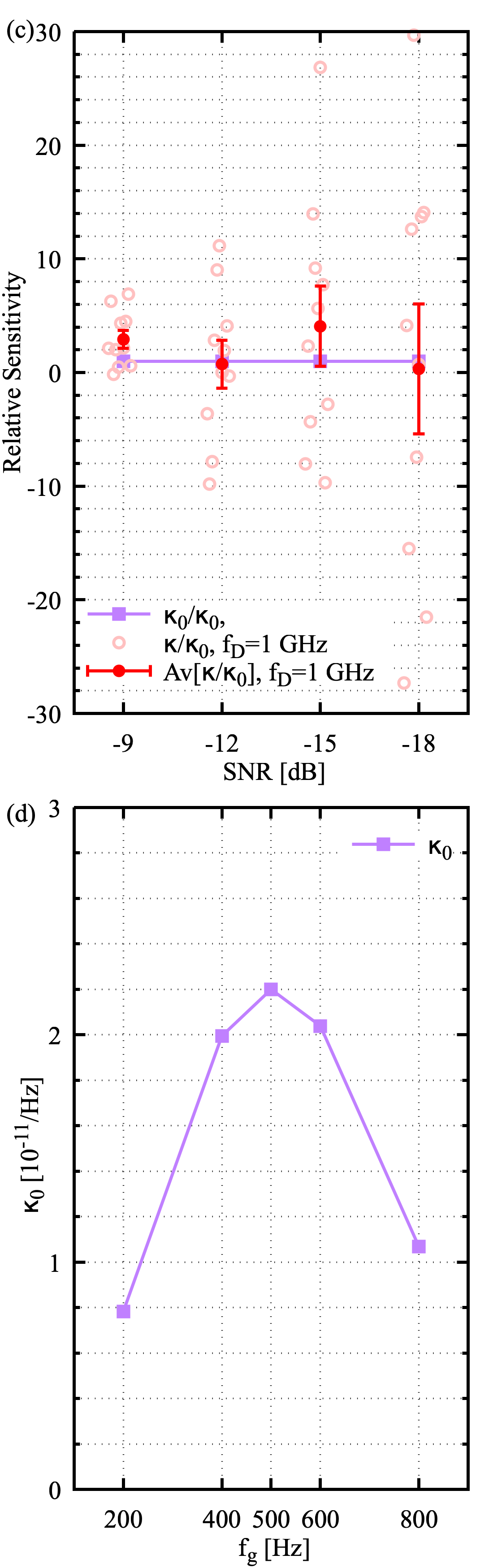}}
	\end{minipage}
}
\caption{\label{Fig:AWVA_resultACandSensitivity}
The simulation results of the sensitivities for GW detection with a strain \red{$h_g =$ $4 \times 10^{-25}$/$\rm \sqrt{Hz}$}. 
(a) The dependency of sensitivities $\kappa(f_g)$ and (b) relative sensitivity $\kappa(f_g)/ \kappa_{0}(f_g)$ on GW frequencies $f_g$, considering various sampling frequencies $f_D$ and SNR = -12 dB.
(c) The dependency of relative sensitivity $\kappa(f_g)/ \kappa_{0}(f_g)$ for detecting a \red{500}-Hz GW on SNR.
(d) The sensitivities of $\kappa_0(f_g)$ \red{for} detecting GWs without noises.
Function Av[ ] represents the mean value of ten measurements and its standard deviation is displayed as the error bar. 
}
\end{figure*}

Fig.~\ref{Fig:AWVA_coffience_shiftCompaaring}(a) presents the results of ${\rm \Theta}_{N}(t)$ \red{for} different GW frequencies $f_g$ at SNR = -12 dB and \red{a sampling frequency of} $f_D$ = 1 GHz. 
\red{In princple}, different $f_g$ values lead to \red{varying ${\rm \Theta}_{N}(t)$ curves}. \red{However,} it is difficult to directly distinguish \red{these} curves with different $f_g$ due to the influence of noises. Therefore, we calculate the sensitivity $\kappa(f_g)$ as well as the relative sensitivity $\kappa(f_g)/ \kappa_{0}(f_g)$ to highlight the dependency of ${\rm \Theta}_{N}(t)$ on $f_g$. Where $\kappa_{0}(f_g)$ represents the sensitivity of the measurements without noises (\red{see Table~\ref{Table:sensitivityDifferentfgwithoutNoise}}). The values of $\kappa(f_g)$ and $\kappa(f_g)/ \kappa_{0}(f_g)$ with various setups are shown in Table~\ref{Table:sensitivityDifferentfg}, Table~\ref{Table:sensitivityDifferentfgfD10GHz}, Table~\ref{Table:sensitivityDifferentfgfD100GHz},Table~\ref{Table:sensitivityDifferentSNR} and Fig.~\ref{Fig:AWVA_resultACandSensitivity}.

Fig.~\ref{Fig:AWVA_coffience_shiftCompaaring}(b) shows the results of ${\rm \Theta}_{N}(t)$ \red{for} different SNRs at the GW frequency $f_{g, max}$ = \red{500} Hz and a sampling frequency of $f_D$ = 1 GHz. The results indicate that as the SNR decreases, it becomes more challenging to distinguish changes in ${\rm \Theta}_{N}(t)$.
However, it can be found in Figs.~\ref{Fig:AWVA_coffience_shiftCompaaring}(c) and \ref{Fig:AWVA_coffience_shiftCompaaring}(d) that increasing the sampling frequency $f_D$ of PDs can further enhance the precision. Note that the sampling frequency $f_D$ is not infinite and is limited  by the performance of PDs.
\red{
Until now, there has been a growing number of reports on high-speed photodetectors with a bandwidth of 100 GHz, driven by the increasing bandwidth demands in optical networks~\cite{Vivien:12,7519081,Salamin2018}.
Therefore, in our simulation, we consider a sampling frequency of 100 GHz as the upper limit, taking into account the capabilities of these high-speed photodetectors.
}

Fig.~\ref{Fig:AWVA_resultACandSensitivity} \red{shows the simulation results of the sensitivity $\kappa(f_g)$ and the relative sensitivity $\kappa(f_g) / \kappa_0(f_g)$ for the GW detection with \red{a} strain $h_g =$ $4 \times 10^{-25}$/$\rm \sqrt{Hz}$}. 
\red{
Fig.~\ref{Fig:AWVA_resultACandSensitivity}(a) demonstrates that the mean values of $\kappa(f_g)$ at sampling frequencies of $f_D = 1$ GHz and $f_D = 10$ GHz are in qualitative agreement with the theoretical value $\kappa_0(f_g)$, except when the GW frequency $f_g$ is equal to 200 Hz. 
However, the error bars (corresponding to the standard deviation) tend to decrease as the sampling frequency $f_D$ increases.
When the sampling frequency can acquire data with a bandwidth of 100 GHz,  the bottom of Fig.~\ref{Fig:AWVA_resultACandSensitivity}(a) indicates that our scheme has sufficient sensitivities to detect GWs in the frequency range of 200 Hz $\leq$ $f_g$ $\leq$ 800 Hz.
More intuitively, Fig.~\ref{Fig:AWVA_resultACandSensitivity}(b) displays the relative sensitivity $\kappa/ \kappa_{0}$, which reflects the robustness against noise.
A smaller deviation of data points from the theoretical value (close to one) indicates better robustness.
Fig.~\ref{Fig:AWVA_resultACandSensitivity}(b) also reveals that measurements near $f_{g} = 500$ Hz, where the peak response occurs, exhibit better robustness.
Furthermore, increasing the sampling frequency can reduce error bars and enhance the precision in AWVA. Fig.~\ref{Fig:AWVA_resultACandSensitivity}(d) displays the theoretical sensitivity dependency on GW frequency without noise, which can potentially be achieved by further increasing the sampling rate.
}

The lowest frequency limit $f_{g, min}$ for GW detection \red{is determined by } the resolution and the sampling frequency of PDs, as well as the size (corresponding to the storage times $\tau_s$) of the zero-area Sagnac interferometer. 
\red{
As shown in Fig.~\ref{Fig:AWVA_resultACandSensitivity}(a),
increasing the sampling frequency can expand the frequency band for GW detections.
With a sampling frequency of $f_D = 100$ GHz, GW measurements at $f_g = 200$ Hz can be effectively performed by measuring $\Theta_N$ with sufficient sensitivity and resolution to distinguish it from other frequency bands.
}
Note that the vertical resolution of PDs is assumed to be capable of detecting such variation in this paper.
In addition, Eq.~(\ref{Eq:gravity_phase_difference}) \red{and Fig.~\ref{Fig:SagnacPhaseAndNoise}(a) demonstrate} that increasing the storage times $\tau_s$ can decrease $f_{g, max}$ of the first peak response as well as $f_{g, min}$.
Therefore, our protocol enables the detection of GWs at $f_{g, \min} = 200$ Hz, and further improvements in PD performance and adjustments to the number of bounces in the zero-area Sagnac interferometer can facilitate the measurement of GWs with even lower $f_g$.

In addition, we investigate the impact  of different SNRs on the measurement of ${\rm \Theta}_{0}(t=t_0)$ in Fig.~\ref{Fig:AWVA_resultACandSensitivity}(c). The results indicate that the error bar gets larger as the SNR increases. 
However, the mean value Av[$\kappa/\kappa{0}$] remains relatively close to the theoretical value of one, indicating the robustness of the measurement against Gaussian noise.
Indeed, this robustness depends on the number $M$ of repeated measurements and the \red{characteristics} of noises. \red{Naturally,} the number of measurements $M$ cannot be infinite. \red{So} it is essential to ensure that $M \times T_A = M \times 6$ $\mu$s is much smaller than the period of the GW (1/$f_g$).
In this paper, we set $M$ to 10, and the relationship of $M \times T_A= 60 $ \redRR{$\upmu$s} $\ll$ 1/(200 Hz) = 0.005 s holds for GW detection in the range of {200} Hz $\leq$ $f_g$ $\leq$ \red{800} Hz.
In general, AWVA \red{demonstrates promising potential for} GW detection with a strain \red{$h_g =$ $4 \times 10^{-25}$/$\rm \sqrt{Hz}$ even in the presence of} Gaussian noises. And it is crucial to extend the investigation beyond stationary Gaussian noise and explore the influence of non-Gaussian noise on AWVA for GW detection in future studies~\cite{Abbott2020-2,Abbott_2020,Jaranowski2012}.

\section{Conclusion} 
In conclusion, we proposed a potential third-generation GW detector based on AWVA, capable of detecting GWs with a strain of \red{$h_g = 4 \times 10^{-25}/\sqrt{\rm Hz}$} in the presence of Gaussian noise. 
\red{Our proposed detector employed a zero-area Sagnac interferometer,}
where a GW with ``plus'' polarization induces a phase difference $\Delta \phi$ in the optical arm. Then the phase difference was amplified as the parameter of post-selection in a weak measurement, \red{while} another simultaneous weak measurement without \red{GW-induced} phase difference was \red{conducted to obtain} the autocorrelation coefficient $\Theta$ of the two pointers.
We simulated AWVA measurements to detect GWs in the frequency range \red{200 Hz $\leq f_g \leq$ 800 Hz}, considering different SNRs and \red{three} sampling frequencies, $f_D$. 
The \red{size of} zero-area Sagnac interferometer \red{was chosen based on the design of the third-generation Einstein Telescope GW detector~\cite{Maggiore_2020}.}
Our results demonstrated that AWVA achieved a measurable sensitivity of \red{the} autocorrelation coefficient $\Theta$. The peak response of $\Theta(f_g)$ occurred at $f_{g, max}$ = \red{500 Hz, which coincided with the target band range of current projects for third-generation GW detectors.} By appropriately setting the size of the interferometer, the peak response frequency can be tuned to the desired GW band.
In addition, our results indicate that increasing the sampling frequency of PDs can enhance the precision of $\Theta$ measurements. The advantage of robustness in AWVA makes it possible to resist noises.

Indeed, our proposed protocol relies on certain limits of existing facilities, \red{such as}  a peak laser power of 1 GW, a detector with a sampling frequency of 100 GHz, and a low resolvable power of 10 pW. \red{Similarly}, other advanced GW detectors also \red{require} high-power lasers to improve the shot-noise limited sensitivity~\cite{RevModPhys.86.121,Willke_2008,Fricke_2012}. Furthermore, several of the key technique can be used to improve AWVA for the GW detection: frequency-dependent squeezed light injections~\cite{PhysRevLett.104.251102,PhysRevLett.129.031101}, a true quantum nondemolition observable readout of acceleration measurement in the GW detection~\cite{PhysRevD.67.122004,Li_2022} and the application of using higher-order spatial modes of the cavity~\cite{PhysRevD.84.102001}.

\ack{
This study is supported by the National Science Foundation of China (Grants No. 42220104002), 
J-H. Huang acknowledges support from the China Scholarship Council (Grants No. 202206410063)}.
\\
\\
\textbf{Data availability statement}\\
\\
All data supporting this study's findings are included in the article.

\appendix
\section{Characteristic results of the AWVA simulation}
\red{We utilized Simulink on MATLAB to conduct simulations of the AWVA measurements for detecting GWs within the frequency range of 200 Hz $\leq f_g \leq$ 800 Hz. Various SNRs and three sampling frequencies were considered in our simulations. The simulation procedures were derived from the methodology described in Reference~\cite{AWVA}. We provide the values of the autocorrelation coefficient $\Theta$, the sensitivity $\kappa(f_g)$, and the relative sensitivity $\kappa(f_g) / \kappa_0(f_g)$ for each measurement. Additionally, we calculated the mean value and error bar based on ten measurements, which are represented by the notation Av[...]. 
Specifically, Table~\ref{Table:sensitivityDifferentfgwithoutNoise} displays the characteristic results obtained from measuring ${\rm \Theta}{N}(t=t_0)$ at different $f_g$ in the absence of noise. 
Table~\ref{Table:sensitivityDifferentfg} shows characteristic results for measuring ${\rm \Theta}_{N}(t=t_0)$ with different $f_g$ at SNR= -12 dB and $f_D$= 1 GHz.
Table~\ref{Table:sensitivityDifferentfgfD10GHz} shows characteristic results for measuring ${\rm \Theta}_{N}(t=t_0)$ with different $f_g$ at SNR= -12 dB and $f_D$= 10 GHz.
Table~\ref{Table:sensitivityDifferentfgfD100GHz} shows characteristic results for measuring ${\rm \Theta}_{N}(t=t_0)$ with different $f_g$ at SNR= -12 dB and $f_D$= 100 GHz.
Table~\ref{Table:sensitivityDifferentSNR} shows characteristic results for measuring ${\rm \Theta}_{N}(t=t_0)$ with different SNRs at $f_{g, max}$= 500 Hz and $f_D$= 1 GHz.
}


\begin{table}[htb!]
\setlength{\tabcolsep}{20.6mm}
\renewcommand{\arraystretch}{0.75}
{
\caption{\label{Table:sensitivityDifferentfgwithoutNoise}
{Characteristic results for measuring ${\rm \Theta}_{N}(t=3\,$ \redRR{$\upmu$s}) $\times 10^{-6}$ at different $f_g$ [Hz] without noises. The units of $\kappa_{0}(f_g)$  is $ 10^{-11}$/Hz.
}}
\vspace*{3mm}
\begin{tabular}{rrr}
\hline
\hline
$f_g$              
& ${\rm \Theta}_{}(t_0)$
& $\kappa_{0}(f_g)$ 
\\ \hline
 000
 &  0.886177789        
 &  -          
 \\
  010
 &  0.886178015        
 &  0.01       
 \\
 200
 &  0.887742946    
 &   0.78
 \\
 400
 & 0.894161373         
 &  1.99      
 \\
\red{500}
 & 0.897180193    
 & 2.20       
 \\
  600
 & 0.898415524         
 &  2.03     
 \\
  800
 & 0.894725321         
 &  1.06      
 \\
\hline
\hline
\end{tabular}
}
\end{table}

\begin{table}[htb!]
\setlength{\tabcolsep}{1.6mm}
\renewcommand{\arraystretch}{0.75}
{
\caption{\label{Table:sensitivityDifferentfg}
{Characteristic results for measuring ${\rm \Theta}_{N}(t=3\,$ \redRR{$\upmu$s}) $\times 10^{-6}$ with different $f_g$ [Hz] at SNR= -12 dB and $f_D$= 1 GHz. The units of $\kappa_{0}(f_g)$  is $ 10^{-11}$/Hz.
}}
\vspace*{1.0mm}
\begin{tabular}{rrrrrrr}
\hline
\hline
$f_g$              
& ${\rm \Theta}_{N}(t_0)$
& $\kappa(f_g)$ 
& $\kappa(f_g)/ \kappa_{0}(f_g)$
& Av[${{\rm \Theta}}_{N}(t_0)$]
& Av[$\kappa(f_g)$]
& Av[$\kappa(f_g)/ \kappa_{0}(f_g)$]
\\ \hline
 \hline
 & 0.890226362          
 & 2.02      
 & 2.58
 \\
 
 & 0.807803072       
 & -39.1          
 & -50.0
 \\
 
 & 0.855157139           
 & -15.5          
 & -19.8
 \\
 
 & 0.820617758  
 & -32.7          
 & -41.8
 \\
 200 
 & 0.812173512          
 & -37.0         
 & -47.2
 & 0.8863
 & 0.38
 &0.49
 \\

 & 0.820829894          
 & -32.6          
 & -41.7
 & ($\pm$ 0.0321)
 & ($\pm$ 16.2)
& ( $\pm$ 20.7)
 \\

 & 1.055421277           
 & 84.6          
 & 108
 \\

 & 0.886129091           
 & -0.02       
 & -0.03
 \\

 & 1.090468402          
 & 102          
 & 130
 \\

 & 0.830656621         
 & -27.7        
 & -35.4
 \\
 \hline
 & 0.835820924         
 & -12.5       
 & -6.31
 \\
 
 & 0.901065362       
 & 3.72        
 & 1.86
 \\
 
 & 0.940735133           
 & 13.6        
 & 6.83
 \\
 
 & 0.849018276  
 & -9.28     
 & -4.65
 \\
 400 
 & 0.815650281          
 & -17.6       
 & -8.83
 & 0.8801
 & -1.51
 & -0.75
 \\

 & 0.873329224         
 & -3.21        
 & -1.61
  & ($\pm$ 0.0147)
  & ($\pm$ 3.67)
  & ($\pm$ 1.84)
 \\

 & 0.861603778          
 & -6.14       
 & -3.07
 \\

 & 0.946102285          
 & 14.9         
 & 7.50
 \\

 & 0.848009767          
 & -9.54         
 & -4.78
 \\

 & 0.929919049        
 & 10.9      
 & 5.48
 \\
 \hline
 & 0.846505965          
 & -7.93     
 & -3.60
 \\
 
 & 0.778020574      
 & -21.6           
 & -9.83
 \\
 
 & 0.800110148         
 & -17.2           
 & -7.82
 \\
 
 & 0.917319271 
 & 6.22         
 & 2.83
 \\
 \red{500}
 & 0.985702502        
 & 19.9         
 & 9.04
 & 0.8944
 & 1.64
 & 0.74
 \\

 & 1.008875716          
 & 24.5        
 & 11.1
   & ($\pm$ 0.0231)
   & ($\pm$ 4.63)
   & ($\pm$ 2.10)
 \\

 & 0.886159446           
 & -0.00       
 & -0.00
 \\

 & 0.907570937         
 & 4.27         
 & 1.94
 \\

 & 0.931289013          
 & 9.02       
 & 4.10
 \\

 & 0.882591427       
 & -0.71     
 & -0.32
 \\
 \hline
 & 0.839906477          
 & -7.71     
 & -3.78
 \\
 
 & 0.905210837      
 & 3.17          
 & 1.55
 \\
 
 & 0.944916739         
 & 9.78      
 & 4.79
 \\
 
 & 0.853309757 
 & -5.47       
 & -2.68
 \\
 \red{600}
 & 0.820091941       
 & -11.0    
 & -5.39
 & 0.8843
 & -0.30
 & -0.14
 \\

 & 0.877386959          
 & 24.5        
 & -0.71
   & ($\pm$ 0.0146)
   & ($\pm$ 2.44)
   & ($\pm$ 1.20)
 \\

 & 0.865922123           
 & -3.37       
 & -1.65
 \\

 & 0.950317457         
 & 10.6      
 & 5.24
 \\

 & 0.852320265          
 & -5.64  
 & -2.76
 \\

 & 0.934239631       
 & 8.01     
 & 3.92
 \\
 \hline
 & 0.836362452          
 & -6.22    
 & -5.83
 \\
 
 & 0.901614925      
 &  1.92           
 &  1.80
 \\
 
 & 0.941289472         
 &  6.88         
 &  6.45
 \\
 
 & 0.849587292 
 & -4.57        
 & -4.28
 \\
 \red{800}
 & 0.816238916        
 &-8.74        
 &-8.18
 & 0.8806
 &-0.68
 &-0.64
 \\

 & 0.873867068          
 &-1.53        
 &-1.44
   & ($\pm$ 0.0147)
   & ($\pm$ 1.83)
   & ($\pm$ 1.72)
 \\

 & 0.862176247           
 & -3.00       
 & -2.80
 \\

 & 0.946661033         
 & 7.56         
 & 7.07
 \\

 & 0.848581261          
 & -4.69       
 & -4.40
 \\

 & 0.930491614       
 & 5.53    
 & 5.18
 \\
\hline
\hline
\end{tabular}
}
\end{table}

\begin{table}[htb!]
\setlength{\tabcolsep}{1.6mm}
\renewcommand{\arraystretch}{0.75}
{
\caption{\label{Table:sensitivityDifferentfgfD10GHz}
{Characteristic results for measuring ${\rm \Theta}_{N}(t=3\,$ \redRR{$\upmu$s}) $\times 10^{-6}$ with different $f_g$ [Hz] at SNR= -12 dB and $f_D$= 10 GHz. The units of $\kappa_{0}(f_g)$  is $ 10^{-11}$/Hz.
}}
\vspace*{1.0mm}
\begin{tabular}{rrrrrrr}
\hline
\hline
$f_g$              
& ${\rm \Theta}_{N}(t_0)$
& $\kappa(f_g)$ 
& $\kappa(f_g)/ \kappa_{0}(f_g)$
& Av[${{\rm \Theta}}_{N}(t_0)$]
& Av[$\kappa(f_g)$]
& Av[$\kappa(f_g)/ \kappa_{0}(f_g)$]
\\ \hline
 \hline
 & 0.900990272          
 & 7.40    
 & 9.45
 \\
 
 & 0.917756624       
 & 15.7          
 & 20.1
 \\
 
 & 0.887766365           
 & 0.79         
 & 1.01
 \\
 
 & 0.897427823  
 & 5.62          
 & 7.18
 \\
 200 
 & 0.904079358          
 & 8.95        
 & 11.4
 & 0.89
 & 5.08
 & 6.49
 \\

 & 0.876374300         
 & -4.90    
 & -6.26
 & ($\pm$ 0.0043)
 & ($\pm$ 2.16)
 & ($\pm$ 2.76)
 \\

 & 0.899396572          
 & 6.60        
 & 8.44
 \\

 & 0.873321681           
 & -6.42       
 & -8.20
 \\

 & 0.907737432          
 & 10.7     
 & 13.7
 \\

 & 0.898614635         
 & 6.21        
 & 7.94
 \\
 \hline
 & 0.921821179         
 & 8.91    
 & 4.46
 \\
 
 & 0.895720975       
 & 2.38        
 & 1.19
 \\
 
 & 0.903052803          
 & 4.21         
 & 2.11
 \\
 
 & 0.905309962  
 & 4.78    
 & 2.39
 \\
 400 
 & 0.884894414          
 & -0.32       
 & -0.16
 & 0.8968
 & 2.66
 & 1.33
 \\

 & 0.883505300         
 & -0.66       
 & -0.33
  & ($\pm$ 0.0043)
  & ($\pm$ 1.09)
  & ($\pm$ 0.55)
 \\

 & 0.898040815          
 & 2.96    
 & 1.48
 \\

 & 0.872006202          
 & -3.54   
 & -1.77
 \\

 & 0.899473257          
 & 3.32        
 & 1.66
 \\

 & 0.904411305        
 & 4.55 
 & 2.28
 \\
 \hline
 & 0.910905486          
 & 4.94     
 & 2.24
 \\
 
 & 0.887978173      
 & 0.36           
 & 0.16
 \\
 
 & 0.911009988         
 & 4.96           
 & 2.25
 \\
 
 & 0.901307522 
 & 3.02         
 & 1.37
 \\
 \red{500}
 & 0.923760846        
 & 7.51         
 & 3.41
 & 0.9007
 & 2.91
 & 1.32
 \\

 & 0.907056340          
 & 4.17        
 & 1.89
   & ($\pm$ 0.0046)
   & ($\pm$ 0.93)
   & ($\pm$ 0.42)
 \\

 & 0.870214054           
 & -3.19       
 & -1.45
 \\

 & 0.904071689         
 & 3.57         
 & 1.62
 \\

 & 0.891996651          
 & 1.16       
 & 0.52
 \\

 & 0.899050515       
 & 2.57     
 & 1.17
 \\
 \hline
 & 0.912168579          
 & 4.33  
 & 2.12
 \\
 
 & 0.889228596      
 & 0.50         
 & 0.24
 \\
 
 & 0.912251916         
 & 4.34      
 & 2.13
 \\
 
 & 0.902558250 
 & 2.73       
 & 1.33
 \\
 \red{600}
 & 0.925006129      
 & 6.47    
 & 3.17
 & 0.90197
 & 2.63
 & 1.29
 \\

 & 0.908307559          
 & 3.68    
 & 1.80
   & ($\pm$ 0.0046)
   & ($\pm$ 0.77)
   & ($\pm$ 0.38)
 \\

 & 0.871417567           
 & -2.46      
 & -1.20
 \\

 & 0.905305139         
 & 3.18   
 & 1.56
 \\

 & 0.893217656          
 & 1.17
 & 0.57
 \\

 & 0.900265534       
 & 2.34     
 & 1.15
 \\
 \hline
 & 0.908395246          
 & 2.77  
 & 2.60
 \\
 
 & 0.885493534     
 &  -0.08           
 &  -0.08
 \\
 
 & 0.908541833         
 &  2.79         
 &  2.61
 \\
 
 & 0.898822140 
 & 1.58      
 & 1.47
 \\
 \red{800}
 & 0.921286204        
 & 4.38   
 & 4.10
 & 0.8982
 & 1.51
 & 1.41
 \\

 & 0.904569834          
 & 2.29    
 & 2.15
   & ($\pm$ 0.0046)
   & ($\pm$ 0.58)
   & ($\pm$ 0.54)
 \\

 & 0.867822143           
 & -2.29      
 & -2.14
 \\

 & 0.901620802         
 & 1.93         
 & 1.80
 \\

 & 0.889570362          
 & 0.42  
 & 0.39
 \\

 & 0.896635962       
 & 1.30
 & 1.22
 \\
\hline
\hline
\end{tabular}
}
\end{table}


\begin{table}[htb!]
\setlength{\tabcolsep}{1.6mm}
\renewcommand{\arraystretch}{0.75}
{
\caption{\label{Table:sensitivityDifferentfgfD100GHz}
{Characteristic results for measuring ${\rm \Theta}_{N}(t=3\,$ \redRR{$\upmu$s}) $\times 10^{-6}$ with different $f_g$ [Hz] at SNR= -12 dB and $f_D$= 100 GHz. The units of $\kappa_{0}(f_g)$  is $ 10^{-11}$/Hz.
}}
\vspace*{1.0mm}
\begin{tabular}{rrrrrrr}
\hline
\hline
$f_g$              
& ${\rm \Theta}_{N}(t_0)$
& $\kappa(f_g)$ 
& $\kappa(f_g)/ \kappa_{0}(f_g)$
& Av[${{\rm \Theta}}_{N}(t_0)$]
& Av[$\kappa(f_g)$]
& Av[$\kappa(f_g)/ \kappa_{0}(f_g)$]
\\ \hline
 \hline
 & 0.890741021          
 & 2.28  
 & 2.91
 \\
 
 & 0.882616419       
 & -1.78          
 & -2.27
 \\
 
 & 0.874441113           
 & -5.86      
 & -7.49
 \\
 
 & 0.881462011  
 & -2.35     
 & -3.01
 \\
 200 
 & 0.891540756          
 & 2.68        
 & 3.42
 & 0.8869
 & 0.36
 & 0.46
 \\

 & 0.896353436         
 & 5.08   
 & 6.49
 & ($\pm$ 0.0021)
 & ($\pm$ 1.06)
 & ($\pm$ 1.36)
 \\

 & 0.885345260          
 & -0.41      
 & -0.53
 \\

 & 0.895504715           
 & 4.66      
 & 5.95
 \\

 & 0.884427598          
 & -0.87  
 & -1.11
 \\

 & 0.886655551         
 & 0.23        
 & 0.30
 \\
 \hline
 & 0.893133405         
 & 1.73    
 & 0.87
 \\
 
 & 0.891006352      
 & 1.20        
 & 0.60
 \\
 
 & 0.899723005          
 & 3.38        
 & 1.69
 \\
 
 & 0.899945170  
 & 3.44
 & 1.72
 \\
 400 
 & 0.896724199          
 & 2.63       
 & 1.32
 & 0.8950
 & 2.20
 & 1.10
 \\

 & 0.888135492         
 & 0.48     
 & 0.24
  & ($\pm$ 0.0013)
  & ($\pm$ 0.33)
  & ($\pm$ 0.16)
 \\

 & 0.890413162          
 & 1.05    
 & 0.53
 \\

 & 0.897141370          
 & 2.74
 & 1.37
 \\

 & 0.899572451         
 & 3.34       
 & 1.67
 \\

 & 0.894242225        
 & 2.01
 & 1.01
 \\
 \hline
 & 0.896207341          
 & 2.00     
 & 0.91
 \\
 
 & 0.903337635      
 & 3.43           
 & 1.55
 \\
 
 & 0.888533119         
 & 0.47          
 & 0.21
 \\
 
 & 0.893482251 
 & 1.46         
 & 0.66
 \\
 \red{500}
 & 0.907663855        
 & 4.29       
 & 1.95
 & 0.8955
 & 1.87
 & 0.85
 \\

 & 0.893966001          
 & 1.55        
 & 0.70
   & ($\pm$ 0.0019)
   & ($\pm$ 0.39)
   & ($\pm$ 0.18)
 \\

 & 0.897835934           
 & 2.33       
 & 1.05
 \\

 & 0.894747491         
 & 1.71      
 & 0.77
 \\

 & 0.886708915          
 & 0.10     
 & 0.04
 \\

 & 0.893107178       
 & 1.38  
 & 0.62
 \\
 \hline
 & 0.897437187          
 & 1.87
 & 0.92
 \\
 
 & 0.904572930      
 & 3.06       
 & 1.50
 \\
 
 & 0.889774038         
 & 0.59   
 & 0.29
 \\
 
 & 0.894712827 
 & 1.42      
 & 0.69
 \\
 \red{600}
 & 0.908900495      
 & 3.78  
 & 1.85
 & 0.8967
 & 1.76
 & 0.86
 \\

 & 0.895200596          
 & 1.50 
 & 0.73
   & ($\pm$ 0.0019)
   & ($\pm$ 0.33)
   & ($\pm$ 0.16)
 \\

 & 0.899067839           
 & 2.14  
 & 1.05
 \\

 & 0.895981706         
 & 1.63   
 & 0.80
 \\

 & 0.887943286          
 & 0.29
 & 0.14
 \\

 & 0.894340653       
 & 1.36  
 & 0.66
 \\
 \hline
 & 0.893763343          
 & 0.94
 & 0.87
 \\
 
 & 0.900882661    
 &  1.83           
 &  1.70
 \\
 
 & 0.886067301         
 &  -0.01   
 &  -0.01
 \\
 
 & 0.891036746 
 & 0.60     
 & 0.56
 \\
 \red{800}
 & 0.905206341        
 & 2.37   
 & 2.20
 & 0.8931
 & 0.86
 & 0.80
 \\

 & 0.891512596          
 & 0.66
 & 0.61
   & ($\pm$ 0.0019)
   & ($\pm$ 0.24)
   & ($\pm$ 0.22)
 \\

 & 0.895387923           
 & 1.15      
 & 1.06
 \\

 & 0.892294849         
 & 0.76        
 & 0.70
 \\

 & 0.884256044          
 & -0.24
 & -0.22
 \\

 & 0.890656017       
 & 0.55
 & 0.51
 \\
\hline
\hline
\end{tabular}
}
\end{table}


\begin{table}[htb!]
\setlength{\tabcolsep}{1.6mm}
\renewcommand{\arraystretch}{0.75}
{
\caption{\label{Table:sensitivityDifferentSNR}
{Characteristic results for measuring ${\rm \Theta}_{N}(t_0)$ $\times 10^{-6}$ with different SNRs [dB] at $f_{g, max}$= \red{500} Hz and $f_D$= 1 GHz. The units of $\kappa_{0}(f_g)$  is $ 10^{-11}$/Hz.
}}
\vspace*{3mm}
\begin{tabular}{rrrrrrr}
\hline
\hline
SNR              
& ${\rm \Theta}_{N}(t_0)$
& $\kappa(f_g)$ 
& $\kappa(f_g)/ \kappa_{0}(f_g)$
& Av[${{\rm \Theta}}_{N}(t_0)$]
& Av[$\kappa(f_g)$]
& Av[$\kappa(f_g)/ \kappa_{0}(f_g)$]
\\ \hline
 & 0.909498596          
 & 4.66      
 & 2.12
 \\
 
 & 0.954938909       
 & 13.7          
 & 6.25
 \\
 
 & 0.884296282           
 & -0.37         
 & -0.17
 \\
 
 & 0.908473954  
 & 4.45        
 & 2.02
 \\
 -9
 & 0.891306031          
 & 1.02
 & 0.46
 & 0.9182
 & 6.41
 & 2.91
 \\

 & 0.933904593          
 & 9.54       
 & 4.33
 &($\pm$ 0.0085)
& ($\pm$ 1.71)
& ($\pm$ 0.77)
 \\

 & 0.909832236        
 & 4.73      
 & 2.15
 \\

 & 0.935728869          
 & 9.91      
 & 4.50
 \\

 & 0.962126961          
 & 15.1   
 & 6.90
 \\

 & 0.892516415     
 & 1.26      
 & 0.57
 \\
 \hline
 & 0.846505965          
 & -7.93     
 & -3.60
 \\
 
 & 0.778020574      
 & -21.6           
 & -9.83
 \\
 
 & 0.800110148         
 & -17.2           
 & -7.82
 \\
 
 & 0.917319271 
 & 6.22         
 & 2.83
 \\
 \red{-12}
 & 0.985702502        
 & 19.9         
 & 9.04
 & 0.8944
 & 1.64
 & 0.74
 \\

 & 1.008875716          
 & 24.5       
 & 11.1
   & ($\pm$ 0.0231)
   & ($\pm$ 4.63)
   & ($\pm$ 2.10)
 \\

 & 0.886159446           
 & -0.00       
 & -0.00
 \\

 & 0.907570937         
 & 4.27         
 & 1.94
 \\

 & 0.931289013          
 & 9.02       
 & 4.10
 \\

 & 0.882591427       
 & -0.71     
 & -0.32
 \\
 \hline
 & 0.797659327          
 & -17.7    
 & -8.04
 \\
 
 & 0.911682410      
 & 5.10          
 & 2.31
 \\
 
 & 0.838502408           
 & -9.53      
 & -4.33
 \\
 
 & 1.039502488  
 & 30.6    
 & 13.9
 \\
 -15
 & 0.987228524         
 & 20.2
 & 9.18
 & 0.9310
 & 8.96
 & 4.07
 \\

 & 0.948195466          
 & 12.4        
 & 5.63
 & ($\pm$ 0.0387)
& ($\pm$ 7.74)
& ($\pm$ 3.52)
 \\

 & 1.181338292          
 & 59.0     
 & 26.8
 \\

 & 0.971323130          
 & 17.0      
 & 7.74
 \\

 & 0.779359809          
 & -21.3        
 & -9.71
 \\

 & 0.855405250       
 & -6.15       
 & -2.79
 \\
 \hline
 & 0.585627536          
 & -60.1    
 & -27.3
 \\
 
 & 0.931736597       
 & 9.11          
 & 4.14
 \\
 
 & 0.715814833         
 & -34.0       
 & -15.4
 \\
 
 & 1.025055698  
 & 27.7     
 & 12.6
 \\
 -18
 & 1.212531803         
 & 65.2
 & 29.6
 & 0.8896
 & 0.69
 & 0.31
 \\

 & 0.804319855
 & -16.3        
 & -7.44
 & ($\pm$ 0.0629)
 & ($\pm$ 12.5)
& ($\pm$  5.72)
 \\

 & 0.894190926           
 & 1.60   
 & 0.72
 \\

 & 1.036893739           
 & 30.1        
 & 13.7
 \\

 & 1.041157040          
 & 30.9     
 & 14.0
 \\

 & 0.649232915      
 & -47.3       
 & -21.5
 \\
\hline
\hline
\end{tabular}
}
\end{table}

\clearpage

\section*{References}
\bibliographystyle{iopart-num.bst}
\bibliography{iopart-num}

\providecommand{\newblock}{}
\begin{thebibliography}{10}
\expandafter\ifx\csname url\endcsname\relax
  \def\url#1{{\tt #1}}\fi
\expandafter\ifx\csname urlprefix\endcsname\relax\def\urlprefix{URL }\fi
\providecommand{\eprint}[2][]{\url{#2}}

\bibitem{PhysRevLett.116.061102}
Abbott B~P {\em et~al.\/} (LIGO Scientific Collaboration and Virgo
  Collaboration) 2016 {\em Phys. Rev. Lett.\/} {\bf 116}(6) 061102
  \urlprefix\url{https://link.aps.org/doi/10.1103/PhysRevLett.116.061102}

\bibitem{PhysRevX.9.031040}
Abbott B~P {\em et~al.\/} (LIGO Scientific Collaboration and Virgo
  Collaboration) 2019 {\em Phys. Rev. X\/} {\bf 9}(3) 031040
  \urlprefix\url{https://link.aps.org/doi/10.1103/PhysRevX.9.031040}

\bibitem{Abbott2020-2}
Abbott B~P {\em et~al.\/} 2020 {\em Living Reviews in Relativity\/} {\bf 23}(1)
  3 \urlprefix\url{https://doi.org/10.1007/s41114-020-00026-9}

\bibitem{Abbott_2020}
Abbott B~P {\em et~al.\/} (LIGO Scientific Collaboration and Virgo
  Collaboration) 2020 {\em Classical and Quantum Gravity\/} {\bf 37} 055002
  \urlprefix\url{https://iopscience.iop.org/article/10.1088/1361-6382/ab685e}

\bibitem{PhysRevD.107.023511}
Badger C, Fornal B, Martinovic K, Romero A, Turbang K, Guo H~K, Mariotti A,
  Sakellariadou M, Sevrin A, Yang F~W and Zhao Y 2023 {\em Phys. Rev. D\/} {\bf
  107}(2) 023511
  \urlprefix\url{https://link.aps.org/doi/10.1103/PhysRevD.107.023511}

\bibitem{PhysRevLett.104.251102}
Eberle T, Steinlechner S, Bauchrowitz J, H\"andchen V, Vahlbruch H, Mehmet M,
  M\"uller-Ebhardt H and Schnabel R 2010 {\em Phys. Rev. Lett.\/} {\bf 104}(25)
  251102
  \urlprefix\url{https://link.aps.org/doi/10.1103/PhysRevLett.104.251102}

\bibitem{RevModPhys.94.025001}
Christensen N and Meyer R 2022 {\em Rev. Mod. Phys.\/} {\bf 94}(2) 025001
  \urlprefix\url{https://link.aps.org/doi/10.1103/RevModPhys.94.025001}

\bibitem{PhysRevD.107.022003}
Tanioka S, Vander-Hyde D, Cole G~D, Penn S~D and Ballmer S~W 2023 {\em Phys.
  Rev. D\/} {\bf 107}(2) 022003
  \urlprefix\url{https://link.aps.org/doi/10.1103/PhysRevD.107.022003}

\bibitem{doi:10.1063/5.0117605}
Cooper S~J, Mow-Lowry C~M, Hoyland D, Bryant J, Ubhi A, O’Dell J, Huddart A,
  Aston S and Vecchio A 2023 {\em Rev. Sci. Instrum.\/} {\bf 94} 014502
  \urlprefix\url{https://doi.org/10.1063/5.0117605}

\bibitem{HLuck_2010}
Lück H, Affeldt C, Degallaix J, Freise A, Grote H, Hewitson M, Hild S, Leong
  J, Prijatelj M, Strain K~A, Willke B, Wittel H and Danzmann K 2010 {\em
  Journal of Physics: Conference Series\/} {\bf 228} 012012
  \urlprefix\url{https://dx.doi.org/10.1088/1742-6596/228/1/012012}

\bibitem{Accadia_2011}
Accadia T {\em et~al.\/} (LIGO Scientific Collaboration) 2011 {\em Classical
  and Quantum Gravity\/} {\bf 28} 114002
  \urlprefix\url{https://dx.doi.org/10.1088/0264-9381/28/11/114002}

\bibitem{Somiya_2012}
Somiya K 2012 {\em Classical and Quantum Gravity\/} {\bf 29} 124007
  \urlprefix\url{https://dx.doi.org/10.1088/0264-9381/29/12/124007}

\bibitem{Hild_2011}
Hild S {\em et~al.\/} 2011 {\em Classical and Quantum Gravity\/} {\bf 28}
  094013 \urlprefix\url{https://dx.doi.org/10.1088/0264-9381/28/9/094013}

\bibitem{RevModPhys.86.121}
Adhikari R~X 2014 {\em Rev. Mod. Phys.\/} {\bf 86}(1) 121--151
  \urlprefix\url{https://link.aps.org/doi/10.1103/RevModPhys.86.121}

\bibitem{PhysRevD.97.064031}
Zhao W and Wen L 2018 {\em Phys. Rev. D\/} {\bf 97}(6) 064031
  \urlprefix\url{https://link.aps.org/doi/10.1103/PhysRevD.97.064031}

\bibitem{Hall_2019}
Hall E~D and Evans M 2019 {\em Classical and Quantum Gravity\/} {\bf 36} 225002
  \urlprefix\url{https://dx.doi.org/10.1088/1361-6382/ab41d6}

\bibitem{luck2020design}
L{\"u}ck H {\em et~al.\/} (ET Steering Committee Editorial Team) 2020 Et design
  report update 2020
  \urlprefix\url{https://apps.et-gw.eu/tds/?content=3&r=17245}

\bibitem{Maggiore_2020}
Maggiore M, Broeck C~V~D, Bartolo N, Belgacem E, Bertacca D, Bizouard M~A,
  Branchesi M, Clesse S, Foffa S, García-Bellido J, Grimm S, Harms J, Hinderer
  T, Matarrese S, Palomba C, Peloso M, Ricciardone A and Sakellariadou M 2020
  {\em Journal of Cosmology and Astroparticle Physics\/} {\bf 2020} 050
  \urlprefix\url{https://dx.doi.org/10.1088/1475-7516/2020/03/050}

\bibitem{PhysRevLett.76.3053}
Sun K~X, Fejer M~M, Gustafson E and Byer R~L 1996 {\em Phys. Rev. Lett.\/} {\bf
  76}(17) 3053--3056
  \urlprefix\url{https://link.aps.org/doi/10.1103/PhysRevLett.76.3053}

\bibitem{Grf_2014}
Gräf C, Barr B~W, Bell A~S, Campbell F, Cumming A~V, Danilishin S~L, Gordon
  N~A, Hammond G~D, Hennig J, Houston E~A, Huttner S~H, Jones R~A, Leavey S~S,
  Lück H, Macarthur J, Marwick M, Rigby S, Schilling R, Sorazu B, Spencer A,
  Steinlechner S, Strain K~A and Hild S 2014 {\em Classical and Quantum
  Gravity\/} {\bf 31} 215009
  \urlprefix\url{https://dx.doi.org/10.1088/0264-9381/31/21/215009}

\bibitem{1913L}
Sagnac G 1913 {\em Proceedings of the USSR Academy of Sciences\/} {\bf 157}
  708--710

\bibitem{RevModPhys.86.307}
Dressel J, Malik M, Miatto F~M, Jordan A~N and Boyd R~W 2014 {\em Rev. Mod.
  Phys.\/} {\bf 86}(1) 307--316
  \urlprefix\url{https://link.aps.org/doi/10.1103/RevModPhys.86.307}

\bibitem{Lundeen2011}
Lundeen J~S, Sutherland B, Patel A, Stewart C and Bamber C 2011 {\em Nature\/}
  {\bf 474} 188--191 ISSN 1476-4687
  \urlprefix\url{https://doi.org/10.1038/nature10120}

\bibitem{Turek_2015}
Turek Y, Kobayashi H, Akutsu T, Sun C~P and Shikano Y 2015 {\em New Journal of
  Physics\/} {\bf 17} 083029
  \urlprefix\url{https://dx.doi.org/10.1088/1367-2630/17/8/083029}

\bibitem{Li_2019}
Li G, Chen L~B, Wang T, He Z~H and Song H~S 2019 {\em Journal of Physics B:
  Atomic, Molecular and Optical Physics\/} {\bf 52} 155501
  \urlprefix\url{https://dx.doi.org/10.1088/1361-6455/ab1500}

\bibitem{Huang_2020}
Huang J~Z, Yu Y, Zhao D, Li H and Zeng G 2020 {\em Journal of Physics B:
  Atomic, Molecular and Optical Physics\/} {\bf 53} 215501
  \urlprefix\url{https://dx.doi.org/10.1088/1361-6455/abb1fe}

\bibitem{Mori_2020}
Mori Y, Lee J and Tsutsui I 2019 {\em Journal of Physics B: Atomic, Molecular
  and Optical Physics\/} {\bf 53} 015501
  \urlprefix\url{https://dx.doi.org/10.1088/1361-6455/ab5200}

\bibitem{Li_2021}
Li G and Yang W 2021 {\em Journal of Physics B: Atomic, Molecular and Optical
  Physics\/} {\bf 54} 165503
  \urlprefix\url{https://dx.doi.org/10.1088/1361-6455/ac09c2}

\bibitem{PhysRevA.97.033851}
Li L, Li Y, Zhang Y~L, Yu S, Lu C~Y, Liu N~L, Zhang J and Pan J~W 2018 {\em
  Phys. Rev. A\/} {\bf 97}(3) 033851
  \urlprefix\url{https://link.aps.org/doi/10.1103/PhysRevA.97.033851}

\bibitem{PhysRevA.105.013718}
Huang J~H, He F~F, Duan X~Y, Wang G~J and Hu X~Y 2022 {\em Phys. Rev. A\/} {\bf
  105}(1) 013718
  \urlprefix\url{https://link.aps.org/doi/10.1103/PhysRevA.105.013718}

\bibitem{AWVA}
Huang J~H, Hu X~Y, Dada A~C, Lundeen J~S, Jordan K~M, Chen H and An J 2022 {\em
  Phys. Rev. A\/} {\bf 106}(5) 053704
  \urlprefix\url{https://link.aps.org/doi/10.1103/PhysRevA.106.053704}

\bibitem{Harraz2022}
Harraz S, Cong S and Nieto J~J 2022 {\em EPJ Quantum Technology\/} {\bf 9} 15
  ISSN 2196-0763
  \urlprefix\url{https://doi.org/10.1140/epjqt/s40507-022-00134-1}

\bibitem{AAV}
Aharonov Y, Albert D~Z and Vaidman L 1988 {\em Phys. Rev. Lett.\/} {\bf 60}(14)
  1351--1354
  \urlprefix\url{https://link.aps.org/doi/10.1103/PhysRevLett.60.1351}

\bibitem{doi:10.1063/1.5019816}
Li D, Guan T, Liu F, Yang A, He Y, He Q, Shen Z and Xin M 2018 {\em Applied
  Physics Letters\/} {\bf 112} 213701
  \urlprefix\url{https://doi.org/10.1063/1.5019816}

\bibitem{He:21}
He Y, Luo L, Xie L, Shao J, Liu Y, You J, Ye Y and Zhang Z 2021 {\em Opt.
  Lett.\/} {\bf 46} 4140--4143
  \urlprefix\url{http://www.osapublishing.org/ol/abstract.cfm?URI=ol-46-17-4140}

\bibitem{PhysRevLett.111.033604}
Xu X~Y, Kedem Y, Sun K, Vaidman L, Li C~F and Guo G~C 2013 {\em Phys. Rev.
  Lett.\/} {\bf 111}(3) 033604
  \urlprefix\url{https://link.aps.org/doi/10.1103/PhysRevLett.111.033604}

\bibitem{PhysRevA.103.032212}
Zhu J, Li Z, Liu Y, Ye Y, Ti Q, Zhang Z and Gao F 2021 {\em Phys. Rev. A\/}
  {\bf 103}(3) 032212
  \urlprefix\url{https://link.aps.org/doi/10.1103/PhysRevA.103.032212}

\bibitem{Shaddock:98}
Shaddock D~A, Gray M~B and McClelland D~E 1998 {\em Appl. Opt.\/} {\bf 37}
  7995--8001
  \urlprefix\url{https://opg.optica.org/ao/abstract.cfm?URI=ao-37-34-7995}

\bibitem{Jaranowski2012}
Jaranowski P and Królak A 2012 {\em Living Reviews in Relativity\/} {\bf
  15}(1) 4 ISSN 1433-8351
  \urlprefix\url{https://link.springer.com/article/10.12942/lrr-2012-4}

\bibitem{https://doi.org/10.1002/lpor.200810016}
Ramachandran S, Fini J, Mermelstein M, Nicholson J, Ghalmi S and Yan M 2008
  {\em Laser $\&$ Photonics Reviews\/} {\bf 2} 429--448
  \urlprefix\url{https://onlinelibrary.wiley.com/doi/abs/10.1002/lpor.200810016}

\bibitem{Kwee:12}
Kwee P, Bogan C, Danzmann K, Frede M, Kim H, King P, P\"{o}ld J, Puncken O,
  Savage R~L, Seifert F, Wessels P, Winkelmann L and Willke B 2012 {\em Opt.
  Express\/} {\bf 20} 10617--10634
  \urlprefix\url{https://opg.optica.org/oe/abstract.cfm?URI=oe-20-10-10617}

\bibitem{Bode:20}
Bode N, Meylahn F and Willke B 2020 {\em Opt. Express\/} {\bf 28} 29469--29478
  \urlprefix\url{https://opg.optica.org/oe/abstract.cfm?URI=oe-28-20-29469}

\bibitem{Meijer:17}
Meijer R~A, Stodolna A~S, Eikema K~S~E and Witte S 2017 {\em Opt. Lett.\/} {\bf
  42} 2758--2761
  \urlprefix\url{https://opg.optica.org/ol/abstract.cfm?URI=ol-42-14-2758}

\bibitem{Fang_2016}
Fang C, Huang J~Z, Yu Y, Li Q and Zeng G 2016 {\em Journal of Physics B:
  Atomic, Molecular and Optical Physics\/} {\bf 49} 175501
  \urlprefix\url{https://doi.org/10.1088/0953-4075/49/17/175501}

\bibitem{PhysRevA.85.052110}
Susa Y, Shikano Y and Hosoya A 2012 {\em Phys. Rev. A\/} {\bf 85}(5) 052110
  \urlprefix\url{https://link.aps.org/doi/10.1103/PhysRevA.85.052110}

\bibitem{lee1949application}
Lee Y~W, Cheatham T and Wiesner J~B 1949 {\em Research Laboratory of
  Electronics, Massachusetts Institute of Technology\/}
  \urlprefix\url{https://dspace.mit.edu/bitstream/handle/1721.1/4912/RLE-TR-141-04718321.pdf}

\bibitem{machines9060123}
Liu Y, Liu J and Kennel R 2021 {\em Machines\/} {\bf 9} ISSN 2075-1702
  \urlprefix\url{https://www.mdpi.com/2075-1702/9/6/123}

\bibitem{Liu:22}
Liu Y, Zhang Y, Xu Z, Zhou L, Zou Y, Zhang B and Hu Z 2022 {\em Opt. Express\/}
  {\bf 30} 18966--18977
  \urlprefix\url{https://opg.optica.org/oe/abstract.cfm?URI=oe-30-11-18966}

\bibitem{mackowiak2015nep}
Mackowiak V, Peupelmann J, Ma Y and Gorges A 2015 {\em Thorlabs, Inc\/} {\bf
  56}
  \urlprefix\url{https://www.thorlabschina.cn/images/TabImages/Noise_Equivalent_Power_White_Paper.pdf}

\bibitem{Vivien:12}
Vivien L, Polzer A, Marris-Morini D, Osmond J, Hartmann J~M, Crozat P, Cassan
  E, Kopp C, Zimmermann H and F\'{e}d\'{e}li J~M 2012 {\em Opt. Express\/} {\bf
  20} 1096--1101
  \urlprefix\url{https://opg.optica.org/oe/abstract.cfm?URI=oe-20-2-1096}

\bibitem{7519081}
Chen H, Galili M, Verheyen P, De~Heyn P, Lepage G, De~Coster J, Balakrishnan S,
  Absil P, Oxenlowe L, Van~Campenhout J and Roelkens G 2017 {\em Journal of
  Lightwave Technology\/} {\bf 35} 722--726

\bibitem{Salamin2018}
Salamin Y, Ma P, Baeuerle B, Emboras A, Fedoryshyn Y, Heni W, Cheng B, Josten A
  and Leuthold J 2018 {\em ACS Photonics\/} {\bf 5}(8) 3291--3297 doi:
  10.1021/acsphotonics.8b00525
  \urlprefix\url{https://doi.org/10.1021/acsphotonics.8b00525}

\bibitem{Willke_2008}
Willke B, Danzmann K, Frede M, King P, Kracht D, Kwee P, Puncken O, Savage R~L,
  Schulz B, Seifert F, Veltkamp C, Wagner S, Weßels P and Winkelmann L 2008
  {\em Classical and Quantum Gravity\/} {\bf 25} 114040
  \urlprefix\url{https://dx.doi.org/10.1088/0264-9381/25/11/114040}

\bibitem{Fricke_2012}
Fricke T~T, Smith-Lefebvre N~D, Abbott R, Adhikari R, Dooley K~L, Evans M,
  Fritschel P, Frolov V~V, Kawabe K, Kissel J~S, Slagmolen B~J~J and Waldman
  S~J 2012 {\em Classical and Quantum Gravity\/} {\bf 29} 065005
  \urlprefix\url{https://dx.doi.org/10.1088/0264-9381/29/6/065005}

\bibitem{PhysRevLett.129.031101}
Heinze J, Danzmann K, Willke B and Vahlbruch H 2022 {\em Phys. Rev. Lett.\/}
  {\bf 129}(3) 031101
  \urlprefix\url{https://link.aps.org/doi/10.1103/PhysRevLett.129.031101}

\bibitem{PhysRevD.67.122004}
Chen Y 2003 {\em Phys. Rev. D\/} {\bf 67}(12) 122004
  \urlprefix\url{https://link.aps.org/doi/10.1103/PhysRevD.67.122004}

\bibitem{Li_2022}
Li J and Zhang T 2022 {\em Classical and Quantum Gravity\/} {\bf 39} 055007
  \urlprefix\url{https://dx.doi.org/10.1088/1361-6382/ac4b98}

\bibitem{PhysRevD.84.102001}
Hong T, Miller J, Yamamoto H, Chen Y and Adhikari R 2011 {\em Phys. Rev. D\/}
  {\bf 84}(10) 102001
  \urlprefix\url{https://link.aps.org/doi/10.1103/PhysRevD.84.102001}

\end{thebibliography}

\end{document}